\documentclass[12pt,a4paper]{article}
\usepackage{amsmath,amsfonts,amssymb,amsthm}
\usepackage{graphicx}
\usepackage{geometry}
\usepackage{url}
\usepackage{bm}
\usepackage{CJK}

\usepackage[super,square,compress]{natbib}
\usepackage[bookmarks=false]{hyperref}

\newcommand{\bs}{\boldsymbol}
\newcommand{\var}{\mathrm{var}}
\newcommand{\hw}{\mathrm{HW}}
\newcommand{\op}{o_{\mathbb{P}}}

\DeclareMathOperator*{\argmin}{arg\,min}

\def\tdei{\tilde{\varepsilon }_{i\cdot}}
\def\var{\textnormal{var}}

\def\int{\textnormal{L}}

\def\pk{p_{[k]}}

\def\sumi{\sum_{i=1}^{n}}

\def\nk{n_{[k]}}
\def\nkt{n_{[k]1}}

\def\var{\textnormal{var}}
\def\cov{\textnormal{cov}}
\def\T{{ \mathrm{\scriptscriptstyle T} }}
\def\cov{\mathrm{cov}}
\def\var{\mathrm{var}}

\def\mx{\mathcal{M}_{\bs{x}}}
\def\mc{\mathcal{M}_{\bs{c}}}

\def\S{K}
\def\pk{p_{[k]}}
\def\nk{n_{[k]}}

\def\pik{\pi_{[k]}}

\def\nkt{n_{[k]1}}

\def\htcht{{\hat{\bs{\tau}}_{ \scalebox{0.6}{\textnormal{ht},$\bs{c}$ }}}}
\def\htxhaj{{\hat{\bs{\tau}}_{ \scalebox{0.6}{\textnormal{haj},$\bs{x}$ }}}}
\def\htht{\hat{\tau}_{ \scalebox{0.6}{$\textnormal{ht}$}}}
\def\hthaj{\hat{\tau}_{ \scalebox{0.6}{$\textnormal{haj}$}}}

\def\htX{\hat{\tau}_X}
\def\htkX{\hat{\tau}_{[k]X}}
\def\Mk{M_{[k]}}

\def\skS{\sum_{k=1}^{K}}

\def\T{{ \mathrm{\scriptscriptstyle T} }}

\def\var{\textnormal{var}}
\def\ni{n_k}
\def\sumij{\sum_{k=1}^{K}\sum_{j \in [k]}}
\def\sumi{\sum_{k=1}^{K}}
\def\sumj{\sum_{j \in [k]}}

\def\maxi{\max\limits_{k=1,\ldots,K}}

\def\maxj{\max\limits_{j \in [k]}}

\def\nti{n_{1k}}
\def\nci{n_{0k}}
\def\unadj{\textnormal{unadj}}

\def\tauunadj{\hat \tau_{\unadj}}

\def\S{S}
\def\s{s}
\def\X{X}
\def\x{X}

\def\Sunadj{\sigma_{\unadj}}
\def\sunadj{\hat \sigma_{\unadj}}

\def\argmin{\mathop{\arg\min}}

\usepackage{scalerel}
\newcommand{\ind}{\scalebox{0.6}{\textrm{IATE}}}
\newcommand{\dir}{\scalebox{0.6}{\textrm{DATE}}}
\newcommand{\tot}{\scalebox{0.6}{\textrm{GATE}}}

\newcommand{\ev}{\scalebox{0.6}{\textrm{EV}}}
\def\rhon{\rho_{\scaleobj{0.8}{n}}}
\def\Var{\textnormal{Var}}
\def\ope{\operatorname{E}}

\def\T{{ \mathrm{\scriptscriptstyle T} }}

\begin{document}

\begin{CJK}{UTF8}{gbsn}

\newtheorem{theorem}{定理}[section]
\newtheorem{assumption}{假设}[section]
\newtheorem{definition}{定义}[section]
\newtheorem{lemma}{引理}[section]
\newtheorem{proposition}{性质}[section]
\newtheorem{remark}{注}[section]


\begin{center}
{\Large 基于设计的因果推断理论}
\end{center}

\begin{center}
卢鑫$^{1}$, 付萬嘉$^{1}$, 李弘梓$^{1}$, 于浩洋$^{1}$, 张泓昊$^{1}$, 朱 珂$^{2,3}$, 刘汉中$^{1}$
\end{center}


\begin{center}
    \small (1. 清华大学统计与数据科学系, 北京, 100084;\\ 2. 北卡罗来纳州立大学统计系, 罗利, NC 27695, 美国;\\ 3. 杜克大学生物统计与生物信息学系, 达勒姆, NC 27710, 美国)
\end{center}
\vspace{0.5cm}

\begin{center}
\begin{minipage}{0.9\textwidth}
\small
{\bf 摘要} 因果推断作为统计学与数据科学的重要研究领域，在医学、经济学、教育学及社会科学等多个应用场景中发挥着核心作用。基于设计的因果推断以随机化试验为出发点，强调依托已知的随机化机制进行统计推断，从而在弱模型依赖下实现因果效应的识别与估计。该范式以 Fisher 与 Neyman 的工作为理论基础，已发展出分层随机化、重随机化等多种设计策略，以及 Fisher 随机化检验、Neyman 渐近推断与回归调整等分析方法。近年来，随着高维数据、个体不依从性与网络干扰等复杂情形的出现，基于设计的因果推断在理论与方法上均取得了显著进展。本文系统回顾该领域的最新研究成果，重点讨论协变量平衡的随机化设计、基于设计的统计推断方法及其在高维、不依从与网络干扰情形下的扩展，为未来因果推断的理论发展与实践应用提供系统性综述与研究展望。

{\bf 关键词}  随机化检验; 回归调整; 不依从; 网络干扰; 模型平均
\end{minipage}
\end{center}

\section{引言}

因果推断是统计学与数据科学中的核心议题，在医学临床试验、公共政策评估、教育干预研究等传统领域具有广泛而深远的应用价值。近年来，随着数据密集型应用的迅猛发展，因果推断在科技行业中也扮演着愈发重要的角色。大型互联网公司常规开展 A/B 测试以评估界面设计或算法调整对用户行为的影响；推荐系统与广告系统依赖因果推断优化策略部署；在社交网络平台，干预设计帮助识别用户之间的传播效应。更进一步，因果推断已成为人工智能领域的关键工具之一，广泛用于解释复杂模型的决策过程、提升算法的公平性与稳健性，并为强化学习中的策略评估提供理论保障。因此，构建具备严谨理论基础、灵活适应不同场景的因果推断方法，已成为统计学与人工智能交叉研究的前沿方向。

不同于描述性或预测性分析，因果推断关注的是某一处理或干预对结局变量的潜在影响，这一问题本质上涉及对“反事实”结果的推理，即在观察到某一处理结果的同时，推断若未接受该处理会出现的结局。为解决这一基本问题，学界发展出多种因果推断理论体系，主要包括：基于潜在结果的框架\citep{Neyman1923,Rubin:1974}、基于因果图模型的结构识别方法\citep{wright1921correlation,pearl2009causality}，以及结构方程模型\citep{wright1921correlation,haavelmo1943statistical}。这些理论各自提供了形式化的因果定义与识别条件，但均依赖对处理机制、协变量与结局变量关系的建模与控制。

在众多因果推断策略中，随机化试验被视为识别因果效应的“金标准”。通过在试验设计阶段将处理以随机方式分配给受试对象，可以在理论上消除协变量混杂所引起的偏倚，从而在弱模型假设下实现因果识别。随着应用需求的增长，试验设计方法逐步发展出更为精细的随机化方案，如分层随机化、重随机化、协变量适应性随机化等，以进一步提升协变量平衡程度与推断效率。

围绕随机化形成的一类研究方向称为“基于设计的因果推断”（design-based causal inference），又称基于随机化的因果推断或者有限总体框架。该方法历史悠久，可追溯至 Fisher 和 Neyman 的工作，并在因果推断的理论与实践中日益发挥重要作用\citep{Neyman1923,Fisher1935,freedman2008regression,Lin2013Agnostic,imbens2015causal,athey2018exact,Abadie:2020,bojinov2023design,Zhao-Ding-biometrika}。基于设计的因果推断强调在明确已知的随机化机制下进行统计推断，侧重于设计阶段对协变量进行平衡，减弱对模型正确性的依赖，尤其适用于复杂试验结构和有限样本量的场景。该框架下的推断方法既包括经典的 Fisher 精确检验与 Neyman 渐近推断，也拓展到因果自助法、回归调整、高维估计等新兴方法，理论研究与实践应用均日益丰富。

在现代的随机化试验中，试验者能够在试验前观察到大量的协变量信息。例如，临床试验中研究者可以观测到病人的身高、体重、基因信息等；在百度、抖音等技术公司，研究者可以观测到每个用户大量的行为数据；当存在干扰时还可以观察到个体之间相互联系和影响的网络信息。平衡协变量在基于设计的因果推断中占据核心地位。平衡协变量有两种主要策略：一是在试验设计阶段通过设计更优的处理分配方案来平衡协变量，二是在试验分析阶段通过回归调整等方法处理协变量的不平衡性。


本文围绕“基于设计的因果推断”展开系统性综述，重点介绍试验设计、统计推断与协变量调整等方面的最新研究进展，涵盖低维与高维设定、不依从性与网络干扰等重要主题。文章结构安排如下：第一部分介绍实现协变量平衡的随机化试验设计方法；第二部分概述基于设计的因果推断方法；第三与第四部分分别讨论低维与高维设置下的协变量调整方法；第五部分探讨不依从性问题；第六部分聚焦于网络干扰情形下的设计与推断方法；最后进行总结与展望。

对于非随机的有限总体量$\bs{a} = (\bs{a}_1,\ldots,\bs{a}_n)$，其中$\bs{a}_i \in \mathbb{R}^m$ ($m \geq 1$)，令$\bar{\bs{a}} = (1/n) \sum_{i=1}^{n} \bs{a}_i$ 表示 $\bs{a}$的总体均值，$\bar{\bs{a}}_1 = (1/n_1) \sum_{i=1}^{n}Z_i \bs{a}_i$表示$\bs{a}$在处理组的样本均值，$\bar{\bs{a}}_0 = (1/n_0) \sum_{i=1}^{n} (1 - Z_i) \bs{a}_i$表示$\bs{a}$在对照组的样本均值，$\bs{S}_{\bs{a}}^2=(n-1)^{-1}\sum_{i=1}^n(\bs{a}_i-\bar{\bs{a}}) (\bs{a}_i-\bar{\bs{a}})^\top$表示$\bs{a}$的有限总体方差，$\bs{s}_{\bs{a}(1)}^2=(n_1-1)^{-1}\sum_{i=1}^n Z_i (\bs{a}_i-\bar{\bs{a}}_1)^2$和$\bs{s}_{\bs{a}(0)}^2=(n_0-1)^{-1}\sum_{i=1}^n (1 - Z_i)( \bs{a}_i-\bar{\bs{a}}_0)^2$分别表示$\bs{a}$在处理组和对照组的样本方差。对于非随机的有限总体量$\bs{a} = (\bs{a}_1,\ldots,\bs{a}_n)$和$\bs{b} = (\bs{b}_1,\ldots,\bs{b}_n)$，令$\bs{S}_{\bs{a}\bs{b}}=(n-1)^{-1}\sum_{i=1}^n( \bs{a}_i-\bar{\bs{a}})(\bs{b}_i-\bar{\bs{b}})^\top$表示两个有限总体的协方差。记$\bs{S}_{\bs{a}}^2 = \bs{S}_{\bs{a}\bs{a}}$。当$m=1$,即$\bs{a}_i$是标量时，我们使用$S$和$s$代替$\bs{S}$和$\bs{s}$。

\section{平衡协变量的试验设计方法}
\setcounter{section}{2}
\setcounter{theorem}{0}
\setcounter{assumption}{0}
\setcounter{definition}{0}
\subsection{分层随机化}
在经典的完全随机化试验中，研究者需事先确定处理组和对照组的样本量，并在所有满足该样本量约束的分配方案中等概率随机选择一种。尽管完全随机化在平均意义上能够平衡组间协变量分布，但由于处理分配随机性的存在，在单次实现中仍有可能出现处理组和对照组协变量分布失衡的情况\citep{Fisher:1926,morgan2012rerandomization}。
例如，考虑一个由20人构成的随机化试验，其中10名男性，10名女性。经过一次完全随机化，是有可能将8名男性和2名女性分配到处理组，其他人分配到对照组的。在这种分配下，处理组男性比例远高于女性，使得性别这个变量出现了不平衡。
这种失衡不仅会引入混杂偏倚，降低因果效应的可解释性，还可能导致处理效应估计的效率损失\citep{imbens2015causal}。协变量失衡在实际应用中频繁出现，例如在教育、经济学或小型临床试验中，样本规模受限，一次完全随机化很有可能发生协变量失衡。因此，亟需引入合理机制以主动提升协变量平衡性，同时不牺牲因果推断的有效性。

在设计阶段，可以采用分层随机化、重随机化、分层重随机化等方法平衡协变量。Fisher \cite{Fisher:1926}首次提出分层随机化策略：在设计阶段，根据关键协变量（如性别、年龄段等离散特征）将试验个体划分为若干互斥层，随后在每个层内独立实施完全随机化。在之前20个试验个体的例子中，可以先根据性别分层，然后在男性和女性中分别独立使用完全随机化，即随机将5名男性分配到处理组，5名女性分配到处理组。这样通过分层随机化实现了平衡性别变量的目的。已有研究表明，当分层协变量对结果变量具有解释作用时，适当分层能够提高估计精度\citep{Fisher:1926,imbens2015causal}。实际上，分层随机化在实证研究中应用广泛。根据Lin \citep{Lin:2015}的调查，2014年发表在主要医学期刊上的224项随机化试验中，近70\%使用了分层随机化。

早期关于分层随机化的研究主要集中于两种典型情形：一类是分层数量少但每层规模较大且每层包含多个处理和对照个体\citep{imbens2011experimental}；另一类是分层数量多但每层规模有限，特别是每层仅包含一个处理个体与一个对照个体的配对设计\citep{abadie2008estimation}。Fogarty \cite{Fogarty:2018a}进一步考虑了精细分层设计，即每个层中要么恰有一个处理个体，要么恰有一个对照个体，同时允许各层规模存在差异。关于渐近理论的研究，对于层数固定而层内样本量趋于无穷的情形，完全随机化下的中心极限定理\citep{li2020}可直接推广至分层随机化试验。Liu和Yang \citep{Liu2020}则考虑了更加一般的情形，即允许大的分层和小的分层同时存在且每个层内的处理组个体所占比例可以不同。该论文证明了简单的分层加权均值差估计量仍具有相合性与渐近正态性。
Liu等人\citep{Liu2024}将渐近理论推广到了处理有多个水平的情形，包括分层随机化$2^K$因子试验。此外，
Zhu等人\citep{zhu2025design}进一步拓展了上述工作，不需要每层至少包含两个处理与对照个体，并适用于高维协变量情形。


\subsection{重随机化}
\subsubsection{重随机化的基本思想和理论}
分层随机化可以平衡离散型协变量。而对于连续型协变量，可以使用重随机化。其基本想法是，在试验开始前，通过重复进行随机化、对不满足协变量平衡性要求的处理分配方案进行舍弃，最终得到并接受一个满足要求的分配方案，然后再开展试验。

Morgan和Rubin \citep{morgan2012rerandomization}对重随机化试验的设计和性质进行了较为系统的考察，提出了一般的重随机化框架。他们明确了重随机化有效的充分条件，对一般的重随机化过程给出了定义，并从理论上给出了重随机化后平均处理效应估计量的有限样本性质。重随机化主要由以下6个步骤组成：

(1) 收集协变量数据；

(2) 明确判定协变量平衡性所使用的准则，即给出接受一个分配方案的标准。例如，处理组和对照组协变量均值之间的马氏距离是否小于某个给定的阈值；

(3) 对所有个体进行随机化处理分配，得到一个分配方案；

(4) 根据已收集的协变量信息检查当前方案是否满足步骤(2)中的标准。如果满足，执行步骤(5)；如果不满足，舍弃当前方案并再次执行步骤(3)；

(5) 使用最终从(4)中保留得到的分配方案进行试验；

(6) 使用随机化检验或渐近理论分析试验结果。

重随机化的重点在于，试验者需要明确协变量平衡性的判断标准。这实际上是需要刻画协变量的组间分布差异。在给出随机化方案的接受准则后，试验者需要以“处理分配满足接受准则”这一事件作为条件，进行条件下的估计和推断。
Morgan和Rubin\citep{morgan2012rerandomization}证明了在接受准则所对应的区域具有一定对称性时，处理组和对照组结果变量的均值差在重随机化下仍是平均处理效应的无偏估计。他们研究了基于处理组和对照组协变量均值之间的马氏距离进行重随机化的方案，
并在线性模型假设下得到使用重随机化后均值差估计量方差降低的百分比。
当协变量与结果变量具有相关性时，重随机化能提升平均处理效应的估计精度。

由于重随机化引入了相对复杂的处理分配机制，在小样本下对平均处理效应进行推断难以实现。在大样本下，Li等人\citep{Li2018}对重随机化后平均处理效应均值差估计量的渐近性质进行了研究，为因果效应的估计和推断建立了理论基础。在采用固定的马氏距离接受阈值、协变量维数固定的情形下，他们证明重随机化后平均处理效应均值差估计量仍具有相合性且其渐近分布是正态分布和截断正态分布的卷积，并给出了重随机化后均值差估计量渐近方差相较于完全随机化下的下降比例。此外，他们提出了渐近方差的保守估计，可以用于构造基于重随机化的渐近有效置信区间。这些结论都是基于设计的，不依赖于结果变量的模型假设。Wang和Li \citep{wang2022rerandomization}完善了重随机化的渐近理论，考虑了允许马氏距离接受阈值随样本量增加趋于零、协变量维数趋于无穷的一般情形。他们给出了重随机化后平均处理效应估计量的渐近分布，并证明了在一定正则化条件下，使用马氏距离的重随机化可以达到理论上的最优精度，从而在渐近意义下使用重随机化相较于使用回归调整而言并不会损失精度。

另一方面，已有研究对重随机化的平衡准则进行了多方面的讨论。
在不同协变量重要程度不同的前提下，Morgan和Rubin\citep{Morgan2015}提出了按协变量重要层级进行重随机化的方法，
对不同重要层级的协变量分别设置平衡性准则，允许将更多的计算资源用于平衡更重要的协变量。Lu等人\citep{Lu2023}研究加权欧式距离作为协变量平衡性判断标准\citep{raab2001balance}的情形，允许试验者按照协变量重要性单独制定权重，其相较于使用马氏距离为协变量赋予相同权重的方法具有一定优势。
为了有效地对协变量重要性进行区分，Liu等人\citep{liu2025bayesian}则从贝叶斯视角设计协变量平衡性准则，对协变量异质性进行量化，提出了采用贝叶斯准则的重随机化方法，并分别从理论与数值的角度说明其方法相较传统方法的优越性。Schindl和Branson \citep{schindl2024unified}考虑了更为一般的以协变量均值差二次型为平衡性度量的重随机化方法，给出了该方案下平均处理效应估计量的渐近性质，并对二次型中半正定矩阵的选择方法进行了讨论。他们证明特定欧氏距离对应的二次型具有极小极大最优性，该平衡性度量下平均处理效应的的估计精度始终接近最优值。
此外，重随机化并不局限于完全随机化试验，在其他类型的试验中也有着广泛应用，例如，因子试验\citep{Li2020factorial}，序贯试验\citep{zhou2018sequential}，随机化调查试验\citep{yang2023rejective}，裂区试验\citep{Shi2025}等。

\subsubsection{分层重随机化}

分层随机化和重随机化都是平衡协变量的重要试验设计方法。Rubin建议将两者结合起来使用，这样做的好处有两个方面：一是加快重随机化的计算速度，二是提高因果效应的估计精度。Rubin将这一设计策略总结为“尽可能分层，无法分层的则进行重随机化”。基于这一思想，Johansson 和 Schultzberg \citep{johansson2022rerandomization}在强零假设、每层的处理组和对照组大小相等、倾向性得分（处理组个体所占比例）相同的情况下研究了将分层和重随机化相结合的方法，证明分层重随机化能提高推断和计算效率。Wang等人\cite{Wang2023}进一步丰富了分层重随机化的方法和理论，允许处理组和对照组规模不等、处理效应非可加、各层倾向性得分不同。该论文提出了基于总体马氏距离和分层马氏距离的两种分层重随机化方法，给出了两种方法的适用范围，并研究了相应的渐近理论，说明了分层和重随机化相结合的优势。


\subsubsection{成对交换重随机化}

Morgan 和 Rubin 提出的重随机化策略需要通过重复生成分配直至马氏距离低于指定阈值来筛选“好”分配（即协变量平衡性较好的分配）。另一类方法如 Krieger 等人的贪婪成对交换算法 \citep{krieger2019nearly}，在局部邻域中迭代地优化分配，使平衡指标单调下降。但这些方法在实际运行中仍面临明显挑战：全局重随机化接受率低使得计算复杂度高，贪婪方法易陷入局部最优，且缺乏足够随机性以开展 Fisher 随机化/精确检验。

为了解决重随机化的计算难题，Zhu和Liu \citep{Zhu2023}提出一种新的方法——成对交换重随机化（Pair-Switching Rerandomization，PSRR），旨在提升协变量平衡性的同时，显著降低生成满足约束的分配的计算成本。PSRR 不再依赖于从完全随机分配中反复筛选，而是从任意初始分配出发，随机选取一个处理组和对照组个体，交换其分配状态以构造新的分配。在每次迭代中，如果新分配的马氏距离变小则接受新分配；若马氏距离变大，则以一个依赖于马氏距离差距的概率接受该分配，从而保留一定的探索性。这种设计类似于模拟退火算法或强化学习中的 $\varepsilon$-贪婪策略，确保了方法在有限时间内能够有效探索平衡分配空间。下面具体介绍PSRR算法。

假设试验中有$n$个个体且稳定个体处理值假设（Stable Unit Treatment Value Assumption; SUTVA）成立。在潜在结果框架中，记第 $i$ （$1\leq i \leq n$）个个体的处理指示变量为 $Z_i \in \{0,1\}$，协变量为 $\bs{x}_i \in \mathbb{R}^p$，在处理状态和对照状态下的潜在结果分别为 $Y_i(1)$ 和 $Y_i(0)$。观测到的结果变量记为$Y_i = Z_i Y_i(1) + (1 - Z_i) Y_i(0)$。记处理组大小为 $n_1 = \sum_i Z_i$，对照组大小为 $n_0 = n - n_1$。个体水平的处理效应可以定义为$\tau_i = Y_i(1) - Y_i(0)$。平均处理效应可以定义为$\tau = (1/n) \sum_{i=1}^{n} \tau_i = (1/n) \sum_{i=1}^{n} \{ Y_i(1) - Y_i(0)  \}$。在分配 $\bs{Z} = (Z_1,\dots,Z_n)$ 下，平均处理效应的均值差估计量定义为
$
\hat{\tau} = {n_1}^{-1} \sum_{i:Z_i=1} Y_i - {n_0}^{-1} \sum_{i:Z_i=0} Y_i$。令$\bar{\bs{x}}_1$ 与 $\bar{\bs{x}}_0$ 分别为处理组和对照组的协变量样本均值，$\bar{\bs{x}}=(1/n) \sum_{i=1}^{n} \bs{x}_i$为协变量的有限总体均值。协变量平衡性可以由马氏距离来度量：
$
M(\bs{Z}) = (\bar{\bs{x}}_1 - \bar{\bs{x}}_0)^\top \bs{S}_{\bs{x}}^{-2} (\bar{\bs{x}}_1 - \bar{\bs{x}}_0)
$，
其中 $\bs{S}_{\bs{x}}^2 = (n-1)^{-1} \sum_{i=1}^{n} (\bs{x}_i - \bar{\bs{x}} )(\bs{x}_i - \bar{\bs{x}} )^\top $ 是协变量的协方差矩阵。

PSRR 的基本流程如下：从初始分配 $\bs{Z}^{(0)}$ 出发，设当前马氏距离为 $M^{(t)}$。在第 $t$ 步，随机抽取 $(i,j)$，满足 $Z_i=1,\ Z_j=0$，交换 $(i,j)$的处理分配状态从而构造新分配 $\bs{Z}^*$ 并计算其马氏距离 $M^* = M(\bs{Z}^*)$。若 $M^* \leq M^{(t)}$，则接受该交换；否则，以概率 $( {M^{(t)}}/{M^*} )^\gamma$ 接受该处理分配状态转移。重复上述步骤，直到 $M^{(t)} \leq a$ 为止，其中 $\gamma > 0$ 控制探索强度，$a$ 为事先设定的马氏距离阈值。

\begin{theorem}[{\cite{Zhu2023}}]
\label{thm:unbiased}
假设 $n_1 = n_0 = n/2$ 且 $\bs{Z}$ 由成对交换重随机化生成，则有 $\mathbb{E}( \hat{\tau} ) = \tau$。相比完全随机化，PSRR 可实现的方差缩减下界为 $(1 - a/p) R^2$，其中$R^2$为协变量与潜在结果之间的多重相关系数的平方，代表协变量解释潜在结果的能力。
\end{theorem}

定理\ref{thm:unbiased}表明在处理组与对照组样本量相等时，$\hat{\tau}$ 是无偏估计量。此外，与完全随机化相比，PSRR 可以降低$\hat{\tau}$的方差，且改进幅度与协变量预测力和阈值设置密切相关。

在很多临床试验的场景中，个体是按先后顺序进入试验的。针对这种情形，Zhu和Liu \citep{Zhu2023}进一步提出序贯成对交换重随机化方法，同时证明了$ \hat{\tau} $在该方法下的无偏性并得到了其方差降低的比例。

\section{基于设计的因果推断方法}
\setcounter{section}{3}
\setcounter{theorem}{0}
\setcounter{assumption}{0}
\setcounter{definition}{0}

在基于设计的因果推断框架下，主要有两种推断方法：一是Fisher随机化检验或精确检验，二是Neyman渐近推断。下面的两小节将分别对此进行介绍。

\subsection{Fisher精确检验}

Fisher精确检验（Fisher exact test），又称 Fisher随机化检验，其思想可以追溯到Fisher 在1935年撰写的《The Design of Experiments》一书中的开创性工作。Fisher认识到，随机化不仅仅是一种控制混淆变量的技术手段，更可以被视为统计推断的基础。他提出，通过随机分配处理状态，我们可以创造一个“人造的概率宇宙”，在这个宇宙中，所有的不确定性都来源于已知的随机化过程，而非未知的抽样变异性或模型误差。

Fisher精确检验的哲学基础体现了一种独特的科学认识论。相比于依赖模型假设的传统方法，Fisher的方法基于实际可操作的随机化过程。Fisher认为，科学推断的可靠性应当建立在可控制和可重复的基础上，而非抽象的概率模型假设上。这一思想对后来的因果推断理论发展产生了深远影响，成为基于设计的因果推断方法的哲学基石。

数学上，Fisher精确检验考虑检验强零假设（Sharp Null）$H_0: Y_i(1) = Y_i(0)$。强零假设所有个体的处理效应为零。这个假设看似过于严格，但它具有重要的方法论价值。在强零假设下，我们可以完全确定每个个体在不同处理状态下的潜在结果，从而能够计算出任何检验统计量在所有可能随机化分配下的精确分布。这种精确性是传统渐近检验方法无法达到的，特别是在小样本情况下具有重要优势。在这个假设下，对于任何检验统计量，我们可以通过枚举所有可能的处理分配来计算其精确分布，从而进行假设检验。这种方法的精确性和非参数性质使得它在现代因果推断中仍然具有重要地位。

Fisher精确检验的核心思想是将观察到的检验统计量与其在强零假设下的精确分布进行比较。具体来说，假设我们观察到的检验统计量为$T_{\mathrm{obs}}$，则强零假设下的精确$p$值（$p$-value）定义为：$p\textnormal{-value} = ({1}/{|\Omega|})\sum_{\bs{z}\in\Omega}I\{T(\bs{z}) \geq T_{\mathrm{obs}}\}$。
其中$\Omega$表示所有可能的处理分配集合，$|\Omega|$表示$\Omega$中的元素个数，$T$为检验统计量，$I(\cdot)$为指示函数，$\bs{z} = (z_1,\ldots,z_n)$为处理分配变量的一个实现。

Fisher精确检验在有限样本下能够准确控制第一类错误率，而不依赖于任何分布假设。因此，Fisher精确检验在小样本和复杂试验设计中具有重要的应用价值。Fisher精确检验的另一个显著特点是其对检验统计量选择的灵活性。理论上，任何依赖于观察数据的统计量都可以作为检验统计量，包括均值差、秩和统计量或更复杂的函数形式。这种灵活性使得Fisher精确检验能够适应各种研究问题和数据特征。

然而，传统Fisher精确检验也存在一些局限性。首先，研究者有可能关心强零以外的原假设，例如，考虑异质性因果效应\citep{ding2016randomization}，分位数因果效应\citep{caughey2023randomisation}，弱零假设(Weak Null，即平均处理效应为0）\citep{wu2021randomization}等。
第二，当样本量较大时，枚举所有可能的处理分配在计算上不可行，因此需要使用Monte Carlo方法来近似。Luo等人 \citep{luo2021leveraging}给出了Monte Carlo近似误差的集中不等式。
第三，虽然Fisher精确检验能够提供$p$值，在实际应用中人们可能关心效应大小的区间，这依赖于对Fisher精确检验的反解\citep{imbens2015causal}。Luo等人 \citep{luo2021leveraging}提出使用二分法来求解，而Zhu和Liu \citep{Zhu2023,zhu2024rejoinder}则提出了显示求解基于随机化的置信区间的方法。

\subsection{Neyman渐近推断}

前面提到，Neyman\citep{Neyman1923}首次系统提出了基于设计的推断框架，这一工作为现代因果推断奠定了重要的理论基础。在这篇文章中，Neyman假设随机化试验中的不确定性仅来源于随机分配，从而提出了基于设计的方差估计量，这个估计量不依赖于任何分布假设，完全基于观察到的数据和已知的随机化机制。这种方差估计方法的提出，为后续的置信区间构造和假设检验奠定了基础。

进入二十一世纪以来，随着统计理论的不断发展和实证研究方法的日趋成熟，研究者们开始深刻认识到基于设计的方法在因果推断中的独特优势和广阔应用前景。这促使学者们开始系统地发展和完善基于设计的渐近推断理论框架，为大样本情况下的因果推断提供了更为可靠的理论基础。在基于设计的渐近推断框架中，关键的理论贡献在于建立了在随机化机制下的中心极限定理。与传统的独立同分布假设下的中心极限定理不同，基于设计的中心极限定理考虑的是固定总体下随机分配的渐近性质。
下面的定理\ref{theorem.nyman}给出了Neyman在完全随机化试验中提出的保守方差估计量，而定理\ref{theorem.asymptotic}则给出了完全随机化试验下基于设计的渐近推断的中心极限定理，这两个定理使得完全随机化试验下有效置信区间的构造成为可能。


\begin{theorem}[\citep{Neyman1923}]
    \label{theorem.nyman}
    在完全随机化试验下，(1) $\hat \tau$是一个无偏估计量：$E(\hat \tau)=\tau$；
(2) $\hat \tau$的方差为：$ \mathrm{var}(\hat \tau) = {S_{Y(1)}^2}/{n_1} + {S_{Y(0)}^2}/{n_0} - {S_{\tau}^2}/{n}$，其中$S_{\tau}^2 = (n-1)^{-1} \sum_{i=1}^{n} (\tau_i - \tau)^2 $为个体处理效应$\tau_i = Y_i(1) - Y_i(0)$的有限总体方差；(3) $\hat \tau$的方差估计量为：$\widehat{\mathrm{var}}_N(\hat \tau)={s_{Y(1)}^2}/{n_1} + {s_{Y(0)}^2}/{n_0}$。这个方差估计量是保守的，即$E\left\{\widehat{\mathrm{var}}_N(\hat \tau)\right\}-\mathrm{var}(\hat \tau) = {S_{\tau}^2}/{n}\geq 0$。其中，等号成立当且仅当$\tau_i=\tau$对所有个体$i=1,\ldots,n$成立，即个体处理效应是常数。
\end{theorem}

\begin{theorem}[\citep{li2017}]
    \label{theorem.asymptotic}
    在完全随机化试验下，如果$S_{Y(1)}^2$，$S_{Y(0)}^2$和$S_{\tau}^2$均有有限的极限，且$\lim_{n\to\infty}n_1/n=C\in(0,1)$，同时$Y_i(z)$满足$(1/n)\max_{z=0,1}\max_{1\leq i\leq n}\left\{Y_i(z)-\bar Y(z)\right\}^2\to 0$, $z=0,1$,则有${ (\hat \tau-\tau )}/{\sqrt{\mathrm{var}(\hat \tau)}}\xrightarrow{d}\mathcal N(0,1)$。
\end{theorem}

除了完全随机化试验，Neyman渐近推断理论还可以扩展到其他类型的随机化试验，例如分层随机化试验\citep{Liu2020}、群组随机化试验\citep{Su2021}和重随机化试验\citep{morgan2012rerandomization,Li2018}等。此外，该理论还能够适应更复杂的情境，包括低维回归调整、高维回归调整、不依从性和网络干扰等问题。这些扩展使得基于设计的渐近推断方法能够在更广泛的试验设计场景中得到应用。

\subsection{方差的上确界和因果自助法}

在上一节中，定理\ref{theorem.nyman}所给出的方差估计量往往过于保守，这可能导致置信区间过宽，从而降低统计检验的功效。为了解决这个问题，Aronow等人\citep{aronow2014sharp}提出了一种基于协方差上确界的方差估计。这种方差估计量在保持保守性的同时，能够提供更为精确的方差估计。Yu等人\citep{yu2025sharp}将上述方差估计方法扩展到分层随机化试验中。

然而，模拟结果显示，当原始的潜在结果接近共单调时，这种方差估计方法会明显低估真实方差，尤其在小样本情况下更为显著。因此，研究者们提出了因果自助法\citep{imbens2021causal,yu2025sharp}（Causal Bootstrap），旨在为小样本情况下提供更稳健的置信区间。
因果自助法是一种基于重抽样的非参数方法，其核心思想是通过对随机化试验数据进行重抽样，利用重抽样生成的检验统计量的分布来估计原始试验中检验统计量的分布，从而构造置信区间。

\section{低维回归调整}
\setcounter{section}{4}
\setcounter{theorem}{0}
\setcounter{assumption}{0}
\setcounter{definition}{0}

回归调整是解决协变量不平衡性、提升因果效应估计精度的常用方法。本章讨论协变量维度固定时，各种常见随机化试验下的回归调整方法和理论。令$Y_i \sim 1 + \bs{u}_i$表示$Y_i$关于$\bs{u}_i$，$i=1,\ldots,n$，带有截距项的最小二乘回归（Ordinary Least Squares; OLS）；令$ Y_i \stackrel{w_i}{\sim} 1 + \bs{u}_i$表示$Y_i$关于$\bs{u}_i$ 带有截距项的加权最小二乘（weighted least squares; WLS）回归，其权重为 $w_i$。如果要在回归公式中添加更多的变量，则使用“+”符号或者$\sum$符号。

\subsection{完全随机化试验}
在完全随机化试验中，处理效应的均值差估计量$\hat{\tau}$可通过线性回归 $Y_i \sim 1 + Z_i$ 中 $Z_i$ 的最小二乘估计系数获得。该回归中$Z_i$系数的异方差稳健方差估计\citep{Huber1967,White1980}的表达式为：
\[
\hat{V}_{\hw} = \frac{(n_1-1)s_{Y(1)}^2}{n_1^2} + \frac{(n_0-1)s_{Y(0)}^2}{n_0^2}.
\]
当样本量较大时，$\hat{V}_{\hw}$与Neyman方差估计量$\widehat{\var}_{N}(\hat{\tau})$近似相等。令$q_{1-\alpha/2}$表示标准正态分布的$1-\alpha/2$分位点。值得注意的是，$\hat{\tau}$与$\hat{V}_{\hw}$都可直接通过线性回归输出，但是基于它们的置信区间$(\hat{\tau}\pm {\hat{V}^{1/2}_{\hw}}q_{1-\alpha/2})$的有效性完全依赖于随机化试验的随机性——这意味着，即便线性模型错误指定，基于它们的推断依然有效，此类方法也因此被称为模型辅助推断。接下来我们介绍一些存在协变量时的模型辅助推断方法。

使用协变量的最简单方法是在回归中直接加入协变量$\bs{x}_i$ \citep{Fisher1935}：
$$
Y_i \sim 1 + Z_i + \bs{x}_i.
$$
该方法为称Fisher回归。对应的 $Z_i$ 系数和异方差稳健方差估计记为$(\hat{\tau}_{\textrm{F}},\hat{V}_{\hw,\textrm{F}})$。但是，
Freedman\cite{freedman2008regression}指出在处理组和对照组样本量不相等（$n_1 \ne n_0$）的完全随机化试验中，基于Fisher回归的处理效应估计量可能会损害估计精度。为了解决这一问题，Lin\cite{Lin2013Agnostic}提出在Fisher回归的基础上引入交互项：
$$
Y_i \sim 1 + Z_i + \bs{x}_i + Z_i\bs{x}_i.
$$
其中$\bs{x}_i$需要预先中心化以满足$\sum_{i=1}^{n}\bs{x}_i=\bs{0}$。对应的 $Z_i$ 系数和异方差稳健方差估计记为$(\hat{\tau}_{\textrm{L}},\hat{V}_{\hw,\textrm{L}})$。与简单均值差估计量相比，Lin回归在渐近意义可以提升估计精度，但交互项的引入增加了变量维度，导致其在小样本或中等样本场景下表现欠佳。Lin\cite{Lin2013Agnostic}中也提到了一种不需要加入交互项的WLS方法——ToM（tyranny-of-minority）回归，可以用来解决这一问题：
\[
Y_i \stackrel{w_i}{\sim} 1 + Z_i + \bs{x}_i,\quad w_i = r_{Z_i}^{-2},\quad r_1 = n_1/n, \quad r_0 = n_0 / n.
\]
对应的 $Z_i$ 系数和异方差稳健方差估计记为$(\hat{\tau}_{\textrm{ToM}},\hat{V}_{\hw,\textrm{ToM}})$。该回归在小或中等样本下表现更优，且能在渐近意义上提升估计精度。通过比较多个稳健统计量，Lu 和 Liu \citep{Lu2024}从理论上解释了ToM 回归在小样本场景中优于 Lin 交互回归的原因。

为了研究上述几个回归估计量的渐近性质，我们定义平均处理效应的线性调整估计量类：
$
\{\hat{\tau}(\bs{\beta}):\hat{\tau}(\bs{\beta}) = \hat{\tau} - (\bar{\bs{x}}_1 - \bar{\bs{x}}_0)^\top \bs{\beta},\ \bs{\beta}\in \mathbb{R}^p \}$，
其中最优的线性调整系数$\bs{\beta}^{\textrm{opt}}$定义为使得方差最小的系数：$\bs{\beta}^{\textrm{opt}} = \argmin_{\bs{\beta} \in \mathbb{R}^p} \var(\hat{\tau}(\bs{\beta}))$。定义$\bs{\beta}_{\textrm{F}} = \bs{S}^{-2}_{\bs{x}}(\bs{S}_{\bs{x}Y(1)}+\bs{S}_{\bs{x}Y(0)})$与$\bs{\beta}_z = \bs{S}^{-2}_{\bs{x}}\bs{S}_{\bs{x}Y(z)}$，并令$e_i(z) = Y_i(z) - \bs{x}_i^\top \bs{\beta}_z$，$\tau_{e,i} = e_i(1)-e_i(0)$，$z=0,1$。
下面的定理展示了三种回归调整估计量的渐近正态性以及对应的异方差稳健方差估计的保守性。令$||\cdot||_\infty$表示向量的$\ell_\infty$范数。
\begin{theorem}[\citep{Lu2024}]
    \label{theorem.asymptotic2}
    在完全随机化试验中，如果$\bs{S}_{Y(z)}^2$、$\bs{S}_{Y(z)\bs{x}}$、$S_{\tau}^2$均有有限的极限，$\bs{S}^2_{\bs{x}}$有正定的极限，且$\lim_{n\to\infty}n_1/n=C\in(0,1)$，同时$Y_i(z)$和 $\bs{x}_i$满足
    \begin{equation*}
       \frac{1}{n}\max_{1 \leq i \leq n} \|\bs{x}_i-\bar{\bs{x}}\|_{\infty}^2,\quad \frac 1n\max_{z=0,1}\max_{1\leq i\leq n}\left\{Y_i(z)-\bar Y(z)\right\}^2\to 0, 
    \end{equation*}
    则有：
    
    （1）回归调整估计量的渐近正态性：
    \begin{equation*}
        \frac{\hat \tau_{\textrm{F}}-\tau}{\sqrt{\mathrm{var}(\hat \tau(\bs{\beta}_{\textrm{F}}))}}\xrightarrow{d}\mathcal N(0,1),\quad \frac{\hat \tau_{\textrm{L}}-\tau}{\sqrt{\mathrm{var}(\hat \tau(\bs{\beta}^{\textrm{opt}}))}}\xrightarrow{d}\mathcal N(0,1),\quad \frac{\hat \tau_{\textrm{ToM}}-\tau}{\sqrt{\mathrm{var}(\hat \tau(\bs{\beta}^{\textrm{opt}}))}}\xrightarrow{d}\mathcal N(0,1),
    \end{equation*}

    （2）异方差稳健方差估计的保守性：
    \begin{eqnarray*}
       && n\hat{V}_{\hw,\textrm{F}} - \mathrm{var}(\hat \tau(\bs{\beta}_{\textrm{F}})) = S^2_{\tau} + \op(1),\quad n\hat{V}_{\hw,\textrm{L}} - \mathrm{var}(\hat \tau(\bs{\beta}^{\textrm{opt}})) = S^2_{e(1)-e(0)} + \op(1),\\
       && n\hat{V}_{\hw,\textrm{ToM}} - \mathrm{var}(\hat \tau(\bs{\beta}^{\textrm{opt}})) = S^2_{\tau} + \op(1).
    \end{eqnarray*}
\end{theorem}

\subsection{分层随机化试验和分层重随机化试验}


上文已经提到，分层随机化是平衡协变量的重要方法。除此之外，研究中通常会观测到许多分层变量外的基线协变量。即使设计谨慎，某些基线协变量在处理组和对照组间出现不平衡的概率仍然很高\citep{Fisher:1926, morgan2012rerandomization}。对此，很多文献使用回归调整寻找更高效的估计量。
Fogarty和Colin\citep{Fogarty:2018a, Fogarty:2018b}讨论了回归调整在两种特殊分层随机化试验（精细分层和配对试验）中的优势。



考虑$n$个个体和$K$个层的分层随机化试验，在层内部使用完全随机化，其中每层可以有不同比例的个体分配到处理组。假设第$k$层包含$n_k$个个体, $n_k \geq 2$，其中$\nti $ 个个体分配到处理组，$\nci = n_k - \nti $ 个个体分配到对照组。
平均处理效应定义为
$\tau  = \sumij \tau_{j} /n = \sum_{k=1}^{K} \pk \bar \tau_{[k]}$，
 其中$\pk = n_k/n $，$\bar \tau_{[k]} = \sumj \tau_{j}/n_k$为第$k$层的平均处理效应，对其它量也有类似定义。
定义${\bar Y}_{[k]1}$和${\bar Y}_{[k]0}$为使用第$k$层处理组和对照组的观测值计算的样本均值（对其它量也有类似定义），一个自然的平均处理效应估计量为
\begin{equation}
\label{eq:sr-unadj}
\tauunadj = \sumi \pk \hat \tau_{[k]}= \sumi \pk \{{\bar Y}_{[k]1}-{\bar Y}_{[k]0}\}.
\end{equation}

对有限总体量$a_i(1),a_i(0)$，$i=1,\ldots,n$，其第$k$层内部的总体均值记为$\bar{a}_{[k]}(1)$和$\bar{a}_{[k]}(0)$，总体方差记为$\S^2_{[k]a(1)}$和$\S^2_{[k]a(0)} $，相应的处理组和对照组的样本均值记为$\bar{a}_{[k]1}$和$\bar{a}_{[k]0}$，样本方差记为$\s^2_{[k]a(1)}$和$\s^2_{[k]a(0)} $，$k=1,\ldots,K$。定义
\begin{equation}
\Sunadj^2 = \sumi \pk^2 \Big\{  \frac{ \S^2_{[k]Y(1)} } { \nti } + \frac{ \S^2_{[k]Y(0)} }{ \nci } -  \frac{ \S^2_{[k] \tau} }{ \ni }  \Big\}.\nonumber
\end{equation}
记$\sunadj^2$为对其前两项将总体方差替换成样本方差得到的方差估计。定义$\pik = \nkt/\nk$。为了得到估计量的渐近正态性和方差的保守估计，我们需要如下假设：

\begin{assumption}
\label{as32_1}
存在常数$\pi_{[k], \infty}$和$C \in (0, 0.5)$使得$C < \min_{k= 1,\ldots, K} \pi_{[k], \infty} \leq  \max_{k= 1,\ldots, K} \pi_{[k], \infty} < 1 - C$且$\max_{k=1,\ldots,K} | \pik  - \pi_{[k], \infty} |  \rightarrow 0$。
\end{assumption}

\begin{assumption}
\label{as32_2}
当$n\rightarrow \infty$时，有
\begin{equation}
\frac{1}{n} \maxi \maxj \big\{ Y_{j}(1) - \bar Y_{[k]}(1) \big\}^2  \rightarrow 0, \quad  \frac{1}{n} \maxi \maxj \big\{ Y_{j}(0) - \bar Y_{[k]}(0) \big\}^2  \rightarrow 0. \nonumber
\end{equation}
\end{assumption}

\begin{assumption}
\label{as32_3}
$\sumi \pk  \S^2_{[k]Y(1)}  / \pik $, $ \sumi \pk  \S^2_{[k]Y(0)}  / (1 - \pik ) $ 和 $\sumi \pk   \S_{ [k]\tau}^2 $存在有限的极限，且前两者的极限为正；$\sumi \pk \big\{ \S^2_{[k]Y(1)}  / \pik  + \S^2_{[k]Y(0)}  / (1 - \pik ) - \S_{ [k]\tau}^2 \big\}$的极限为正。
\end{assumption}

\begin{theorem}[\citep{Liu2020}]
\label{th32_1}
若假设\ref{as32_1}--\ref{as32_3}成立，则$( \tauunadj - \tau ) / \Sunadj \xrightarrow{d} \mathcal{N}(0,1)$。此外，如果对于每层$k=1,\ldots,K$,  有$2\leq \nti  \leq n_k - 2$，则
$n \sunadj^2 - n \Sunadj^2 = \sumi \pk \S_{ [k]\tau}^2 + \op(1)$。
\end{theorem}

由于篇幅关系，对于协变量调整，我们仅叙述其原理和主要结论。首先定义其通式：
\begin{equation}
\hat \tau_{\textnormal{ols}, {\bs{\beta}}} = \sumi \pk\bigg\{\Big[  \{{\bar Y}_{[k]1} - \big\{  {\bar{\bs{x}}}_{[k]1}  - {\bar{\bs{x}}}_{[k]} \big\}^\T \hat {\bs{\beta}}_{1k} \Big]  - \Big[ \{{\bar Y}_{[k]0} - \big\{ {\bar{\bs{x}}}_{[k]0}  - {\bar{\bs{x}}}_{[k]} \big\}^\T  \hat {\bs{\beta}}_{0k} \Big]\bigg\}, \nonumber
\end{equation}
其中，$\hat {\bs{\beta}}_{1k} $ 和 $\hat {\bs{\beta}}_{0k} $为估计得到的回归系数。若使用每个处理下的全部个体做加权线性回归，即要求$\hat {\bs{\beta}}_{1k} $ 和 $\hat {\bs{\beta}}_{0k} $与$k$无关，则得到整体策略的回归调整；若逐层做线性回归，即允许$\hat {\bs{\beta}}_{1k} $ 和 $\hat {\bs{\beta}}_{0k} $对每个$k$各不相同，则得到层异质策略。根据文献\citep{Liu2020,Lu2024}，我们有以下主要结果：
\begin{itemize}
\item[（i）] 若协变量$\bs{x}_i$满足平行于假设\ref{as32_2}-\ref{as32_3}的条件，以及$\bs{x}_i$和$Y_i(1),Y_i(0)$的协方差满足类似的非退化条件，则整体策略的回归调整估计量具有渐近正态性，且在$\pi_{[k], \infty}$全部相等时保证相对于不调整的估计量有精度提升；
\item[（ii）] 若随着$n$增大，假设$K$有界，且关于协变量$\bs{x}_i$以及$\bs{x}_i$和$Y_i(1),Y_i(0)$的协方差的某些极限和正则性条件成立，则层异质策略的回归调整估计量具有渐近正态性，且保证相对于不调整的估计量有精度提升。
\end{itemize}

    对于重随机化和分层随机化的结合使用，这里简述其做法和理论性质。Wang等人 \citep{Wang2023}提出了基于总体马氏距离和分层马氏距离的分层重随机化方法：
\begin{itemize}
\item[(i)]基于总体马氏距离的分层重随机化方法。定义
\begin{equation}
\label{eq:sr-x}
\htX=\skS\pk  \{  {\bar{\bs{x}}}_{[k]1} -  {\bar{\bs{x}}}_{[k]0} \}=\skS\pk\htkX
\end{equation}
以及$M_{\htX}=\htX^\T\cov(\htX)^{-1}\htX$。当且仅当$M_{\htX}<a$时接受分层随机化给出的处理分配，其中$a$为预先指定的阈值常数； 

\item[(ii)] 基于分层马氏距离的分层重随机化方法。定义分层马氏距离为
$$\Mk=\htkX^\T\cov(\htkX)^{-1}\htkX,\ k=1,\ldots,K,$$
当且仅当对于所有的$k=1,\ldots,K$,都有$\Mk<a_k$时接受分层随机化给出的处理分配，其中$a_k$为预先指定的阈值常数。
\end{itemize}

    在类似假设\ref{as32_1}-\ref{as32_3}的条件下，第一种分层重随机化方法得到的估计量有极限分布，而第二种分层重随机化方法得到的估计量需要接近“层数有限且每层样本量趋于无穷大”的条件才能保证极限分布存在，但在极限分布存在时相对前者有更高的精度。

在分层重随机化试验下也可以使用回归调整进一步提升估计精度。Lu和Liu \citep{Lu2024}将ToM回归推广到分层随机化和分层重随机化试验中。令$\bs{\delta}_{i} = (I(i\in [2])-\pi_{[2]},\ldots,I(i\in [K])-\pi_{[K]})$为个体$i$中心化后的层指示变量。ToM回归的具体形式为
 \begin{equation*}
   Y_{i} \stackrel{w_i}{\sim} 1 + Z_{i}+\bs{\delta}_{i}
 +Z_{i} \bs{\delta}_{i}+
 \bs{x}_{i},
 \end{equation*}
 其中权重
 $ w_{i}=Z_{i}\pi_{[k]}^{-2} \{ \nti  / (\nti -1) \} +(1-Z_{i})(1-\pi_{[k]})^{-2}\{ \nci / (\nci -1) \}$。相比于Lin的回归，其将协变量与层指示变量和处理指示变量的交互项都加入到回归表达式中的做法，ToM回归只加入协变量，减少了估计参数的维度，具有更好的有限样本表现。
 
\subsection{因子试验}
因子试验被广泛应用于研究多个因子对某一响应的联合效应\citep{wu2011experiments,dasgupta2015causal}。在因子试验中，同样存在平衡协变量的问题。因此，因子试验中经常使用分层随机化方法。
研究者通常使用协方差分析来分析分层随机化（随机化区组）因子试验的结果。该方法需要假设一个包含固定或随机区组效应的线性模型，但随机化本身并不能保证线性模型所需的“常规”假设（如线性、正态性和误差同方差）成立。为了放宽上述线性模型假设，可以考虑基于设计的推断框架。
在完全随机化的 $2^K$因子试验中，Dasgupta 等人 \citep{dasgupta2015causal}使用潜在结果定义了因子效应，并探索了针对强零假设的Fisher精确检验；
Liu等人 \citep{Liu2024}建立了分层随机化因子试验中因子效应估计量的联合分布的渐近理论，并提出协变量调整方法提高因子效应的估计精度。

下面对分层随机化$2^K$因子试验中感兴趣的因子效应进行定义。考虑一个包含$n$个个体和$K$个因子的$2^K$因子试验（$K\geq 1$）。其中，每个因子有两个水平（分别记为$-1$ 和 $+1$），从而共有$Q = 2^K$种处理组合。沿用先前关于分层随机化试验的记号，在随机化前，将这$n$个个体依据某些重要离散变量（如性别、地理位置）进行分层，共分为$B$个层（层下标为 $[m]$），随机化在层间独立进行（为了和因子数$K$区分，这里使用$B$表示层个数）。记 $n_{[m]}$ 为第$m$个层的个体数，满足 $n_{[m]} \geq Q$ 且 $\sum_{m=1}^B n_{[m]} = n$。在层 $m$ ($m=1,\dots,B$) 内，共计有$n_{[m]q}$个个体被随机分配至处理组合 $q$ ($q=1,\dots,Q$)，其中 $n_{[m]q} \geq 1$。处理组合 $q$ 的总样本量为 $n_q = \sum_{m=1}^M n_{[m]q}$。

设 $Y_i(q)$ 为个体 $i$ 在处理组合 $q$ 下的潜在结果，$Y_i$为个体 $i$观测到的结果。
个体水平因子效应可定义为潜在结果的线性对比，但由于试验中每个个体仅接受单一的处理组合，个体水平效应在无额外模型假设下不可识别。在稳定个体处理值假设下，可以估计总体的平均因子效应。
令 $\boldsymbol{\iota}_q = (\iota_{q1},\dots,\iota_{qK})^\top \in \{ -1,+1 \}^K$ 
表示处理组合 $q$ 的因子水平配置，
$\bar{Y}_{[m]}(q) = n_{[m]}^{-1} \sum_{i \in [m]} Y_i(q)$ 
为层 $m$ 处于处理组合$q$时的潜在结果均值，
则层 $m$ 中因子 $k$ 的平均主效应定义为：$\tau_{[m]k} 
= 2^{-(K-1)}  \sum_{q=1}^Q \iota_{qk} \bar{Y}_{[m]}(q) 
= 2^{-(K-1)}  \bs{g}_k^\top \bar{\bs{Y}}_{[m]}$, $k=1,\ldots,K$，
其中，$\bs{g}_k = (\iota_{1k},\dots,\iota_{Qk})^\top$ 
称为因子 $k$ 的生成向量，
$\bar{\bs{Y}}_{[m]} = (\bar{Y}_{[m]}(1), \dots, \bar{Y}_{[m]}(Q))^\top$。 
多个因子之间的交互效应可以通过 $\bm{g}$ 生成向量来定义，它是各个因子主效应对应位置上的元素逐位相乘的结果。更具体地，对于 $1 \le f \le F = 2^K - 1 = Q - 1$，令 $\bm{g}_f = (g_{f,1}, \ldots, g_{f,Q})^\top \in \{-1, +1\}^Q$ 为第 $f$ 个因子主效应或交互效应的生成向量，其满足 $\sum_{q=1}^Q g_{f,q} = 0$。定义$\tau_{[m]f} = 2^{-(K-1)} \bs{g}_f \bar{\bs{Y}}_{[m]}$。我们用一个 $F$ 维列向量 $\bm{\tau}_{[m]} = (\tau_{[m]1}, \ldots, \tau_{[m]F})^\top$ 表示层 $m$ 中所有的平均因子效应。
令$p_{[m]} = n_{[m]}/n$ 表示层 $m$ 的个体占比，$\bm{d}_q = (g_{1,q},\dots,g_{F,q})^\top$。我们有$\bs{\tau}_{[m]} = 2^{-(K-1)}  \sum^Q_{q=1} \bs{d}_q \bar{Y}_{[m]}(q)$，其中，$\sum_{q=1}^Q \bs{d}_q = \bs{0}$。
类似地，令总体的平均潜在结果为 $\bar{Y}(q) = n^{-1} \sum_{i=1}^{n} Y_i(q) = \sum_m p_{[m]} \bar{Y}_{[m]}(q)$，则总体的因子效应向量为：
\begin{align*}
\bs{\tau} = \dfrac{1}{2^{K-1}} \sum^Q_{q=1} \bs{d}_q \bar{Y}(q) = \sum_{m=1}^{B} p_{[m]} \bs{\tau}_{[m]}.
\end{align*}

下面对$\bs{\tau}$进行估计。不考虑额外的协变量信息时，定义$\hat {\bar Y}_{[m]}(q) = (1/n_{[m]q})\sum_{i \in [m]} I(Z_i = q)Y_i$为使用观测值估计的第$m$层处理组合$q$下的样本均值，$\hat {\bar Y}(q) = \sum_{m=1}^{B} p_{[m]} \hat {\bar Y}_{[m]}(q)$（对协变量也有类似定义），以及$ \widehat {\bs{\tau}}_{[m]} = 2^{-(K-1)}  \sum^Q_{q=1} \bs{d}_q \hat {\bar Y}_{[m]}(q)$，可以使用插入法得到$\bs{\tau}$（未协变量调整）的无偏估计：
$$\widehat { \bs{\tau} }_{\unadj} = \sum_{m=1}^B  p_{[m]}  \widehat {\bs{\tau}}_{[m]} =  \dfrac{1}{2^{K-1}}   \sum^Q_{q = 1} \bs{d}_q  \widehat{\bar{Y}}(q).$$
Liu等人 \citep{Liu2024}在一定条件
下证明了$\widehat { \bs{\tau} }_{\unadj}$是相合的和渐近正态的。

沿用前文的记号，对每个个体$i$，我们额外观测到$p$维基线协变量$\X_{i} =  (\x_{i1},\ldots, \x_{ip})^\top$，并用这些协变量进行回归调整以提高对$\bs{\tau}$的估计精度。利用$\widehat{ \bar{Y}} (q) - \big \{ \widehat{\bar{\X}}(q) -  \bar{\X} \big \} ^\T \widehat {\boldsymbol \beta}(q)$代替$\widehat { \bs{\tau} }_{\unadj}$中的$\widehat{\bar{Y}}(q)$进行求和以实现对$\bar{Y}(q)$更精确的估计。基于这一思想，Liu等人\citep{Liu2024}提出了协变量调整后的因子效应估计量：
\begin{align*}
 \widehat {  \bs{\tau} }_{\text{adj}} =  \dfrac{1}{2^{K-1}}   \sum^Q_{q = 1} \bs{d}_q \Big[ \widehat{ \bar{Y}} (q) - \big \{ \widehat{\bar{\X}}(q) -  \bar{\X} \big \} ^\T \widehat {\boldsymbol \beta}(q) \Big].
\end{align*}
其中，调整系数$\widehat {\boldsymbol \beta}(q)$是对${\boldsymbol \beta}(q)  =  \argmin_{ \boldsymbol \beta }  \var \big[  \widehat{ \bar{Y}} (q) - \{ \widehat{\bar{\X}}(q) -  \bar{\X}  \} ^\T \boldsymbol \beta   \big]$
的WLS估计。Liu等人 \citep{Liu2024}证明了协变量调整后$\widehat {  \bs{\tau} }_{\text{adj}}$仍具有渐近正态性，但$\widehat {  \bs{\tau} }_{\text{adj}}$的渐近方差并不是总小于$\widehat { \bs{\tau} }_{\unadj}$。只有在不进行分层（$B=1$）或每个层都具有相同倾向性得分的条件下，上述协变量调整才能保证渐近方差的降低。

在现实应用中，每个层均具有相同倾向性得分这一条件并不总能满足。为了提高各层倾向性得分不同时${ \bs{\tau} }$的估计效率，Liu等人 \citep{Liu2024}从条件推断的角度提出了两种新的协变量调整方法：对第$f$个因子效应${  \tau }_{f}$，条件在其对应的协变量因子效应估计$\widehat {  \tau }_{\X,f}$上单独估计该因子效应,得到对应的估计量$\widehat{  \tau }_{f,\text{cond}}$以及${  \bs{\tau} }$的估计量$ \widehat{\bs{\tau}}_{\text{cond}} =(\widehat{ \tau}_{1,\text{cond}},\dots,\widehat{ \tau}_{F,\text{cond}})^\T$；或对所有因子效应${  \bs{\tau} }$，条件在所有协变量因子效应估计$ \widehat{\bs{\tau}}_{\X} =(\widehat{\boldsymbol \tau}_{\X,1}^\T,\dots,\widehat{\boldsymbol \tau}_{\X, F}^\T)^\T$
上同时估计所有因子效应,得到${  \bs{\tau} }$的估计量$ \widehat{\bs{\tau}}_{\text{cond2}}$。另外，当每个层的样本量均足够大时，Liu等人 \citep{Liu2024}提出了在每个层内部进行协变量调整的方法，并给出了相应的$\bs{\tau}$的估计量$\widehat{\bs{\tau}}_{\text{inter}}$。该文章证明了上述估计量的渐近正态性，并给出了各估计量对应的保守方差估计。

相较于$\widehat { \bs{\tau} }_{\unadj}$，$\widehat{{\bs{\tau}}}_{
\text{cond}}$ 在估计各因子效应 $\tau_f$ 时效率更高，但对某些系数矩阵 ${C}$，估计 ${C} {\bs{\tau}}$ 时效率可能下降。一般地，$\widehat{{\bs{\tau}}}_{\text{cond2}}$ 在估计效率上优于 $\widehat{{\bs{\tau}}}_{\text{cond}}$ 和 $\widehat{{\bs{\tau}}}_{\text{unadj}}$。此外，当各层倾向得分相同时，$\widehat{{\bs{\tau}}}_{\text{cond2}}$ 与 $\widehat{{\bs{\tau}}}_{\text{adj}}$ 渐近等价，但需估计更多调整系数。当层个数固定且每个层的样本量趋于无穷大时，$\widehat{{\bs{\tau}}}_{\text{inter}}$是最优的。因此，Liu等人\citep{Liu2024}建议,在存在小型层且各层倾向得分相同时使用 $\widehat{{\bs{\tau}}}_{\text{adj}}$；存在小型层但倾向得分不相同时使用 $\widehat{{\bs{\tau}}}_{\text{cond2}}$；仅存在大型层时使用 $\widehat{{\bs{\tau}}}_{\text{inter}}$。

\subsection{群组随机化试验和群组重随机化试验}
群组随机化广泛应用于临床试验和政策评估\citep{hayes2017cluster}。其核心特点是，群组作为基本分配单元，同一群组内的所有个体都会接受相同的处理。它主要应用于两类场景：个体之间存在交互作用从而影响因果效应的识别，或者个体水平的随机化试验受到限制。常见的群组包括医院、社区、村庄、学校班级或公司部门等。
举例来说，若要比较两种手术流程（A 与 B）的效果，对每位患者进行随机化比较难于实施，而将每个医院看成整体进行随机化分配（相同医院的患者分配到相同的手术流程）更符合实际情况。此时，可采用群组随机化设计：将参与试验的多家医院随机分配至处理组（采用手术流程 A）或对照组（采用手术流程 B），最终通过比较两组医院内患者的预后指标（如术后恢复时间、并发症发生率等）评估手术流程差异。在此例中，医院即为 “群组”，而接受手术的患者则是“个体”。

相较于完全随机化试验，群组随机化试验更容易出现协变量失衡的问题。以基于医院的群组随机化试验为例，首先，群组随机化试验涉及的协变量通常更多，不仅包括个体层面的协变量（如性别、年龄、生理指标等），也包括群组层面的协变量（如医院的年均手术量和医疗水平等）\citep{hayes2017cluster}。协变量数量的增加本身就提高了协变量失衡的可能性。其次，尽管试验可能涵盖大量个体，但群组的数量往往较少，例如虽然每家医院的患者众多，但参与试验的医院总数通常有限，这进一步加剧了群组层面协变量失衡的风险。为了解决这一问题，我们将介绍设计阶段的重随机化方法以及分析阶段的回归调整方法。

考虑 $n$ 个个体、 $m$ 个群组的群组随机化试验。试验者随机将 $m_1$ 个群组分配到处理组，将 $m_0=m-m_1$ 个群组分配到对照组。记$Z_i$为群组$i$的处理指示变量，当群组被分配到处理组时 $Z_i=1$，被分配到对照组时 $Z_i=0$。记 $n_{(i)}$ 为群组 $i$ 的大小，并有 $\sum_{i=1}^m n_{(i)} = n$。用 $(i,j)$ 或简写为 $ij$ 表示群组 $i$ 中的第 $j$ 个个体。记 $Z_{ij}$ 为个体 $(i,j)$ 的处理指示变量，群组随机化试验满足 $Z_{ij} = Z_{i}$， $j=1,\ldots,n_{(i)}$。用$n_1 = \sum_{i=1}^{m} Z_i n_{(i)}$ 和 $n_0=n-n_1$ 分别表示被分配到处理组和对照组的个体数量。值得注意的是，如果 $n_{(i)}$ 不全相等，$n_1$ 和 $n_0$ 是随机的。令 $Y_{ij}(1)$ 和 $Y_{ij}(0)$ 分别表示个体 $(i,j)$ 在处理和对照下的潜在结果。本章采用有限总体的框架，固定潜在结果，考虑 $Z_i$ 为唯一的随机性来源。由于 $Z_i$ 的随机性，观测结果 $Y_{ij}= Z_iY_{ij}(1)+(1-Z_i)Y_{ij}(0)$ 是随机的。平均处理效应定义为：
$\tau = n^{-1}\sum_{i=1}^{m}\sum_{j=1}^{n_{(i)}} \{Y_{ij}(1)-Y_{ij}(0)\}$。

平均处理效应的常用的估计量包含Hajek估计量和Horvitz--Thompson（HT） 估计量，其定义如下：
\begin{align*}
    & \hat{\tau}_{\textnormal{haj}} = n_1^{-1} \sum_{i=1}^m\sum_{j=1}^{n_{(i)}}Z_{ij}Y_{ij} - n_0^{-1} \sum_{i=1}^m\sum_{j=1}^{n_{(i)}} (1-Z_{ij})Y_{ij},\\
    & \hat{\tau}_{\textnormal{ht}} =  m_1^{-1}\sum_{i=1}^m Z_i\tilde{Y}_{i\cdot}-m_0^{-1}\sum_{i=1}^m (1-Z_i)\tilde{Y}_{i\cdot},
\end{align*}
其中$\tilde{Y}_{i\cdot} = \sum_{j=1}^{n_{(i)}}Y_{ij} m/n$。 $\hat{\tau}_{\textnormal{haj}}$使用个体水平变量$Y_{ij}$的均值差，$\hat{\tau}_{\textnormal{ht}}$使用群组水平变量$\tilde{Y}_{i\cdot}$的均值差。Su和Ding \citep{Su2021} 证明了对于 $\star = \textnormal{ht}, \textnormal{haj}$，有 $m^{1/2} (\hat\tau_{\star} -\tau)/\sigma_{\star} \xrightarrow{d} \mathcal{N}(0, 1)$，其中，
\begin{gather*} \sigma^2_{\textnormal{ht}} =  \frac{S^2_{\tilde{Y}(1)}}{m_1} + \frac{ S^2_{\tilde{Y}(0)}}{m_0} - \frac{S^2_{\tilde{Y}(1)-\tilde{Y}(0)}}{m}, \quad \sigma^2_{\textnormal{haj}} = \frac{S^2_{\tilde{{\varepsilon} } (1)}}{m_1}  + \frac{S^2_{\tilde{{\varepsilon} } (0)}}{m_0}  - \frac{S^2_{\tilde{{\varepsilon} } (1)-\tilde{{\varepsilon} } (0)}}{m}. \end{gather*}
上式中，$\tilde{\varepsilon}_{i\cdot}(z) = \sum_{j=1}^{n_{(i)}}\{Y_{ij}(z)-\bar{Y}(z)\}m/n$， $S^2_{\cdot}$ $(\cdot \in \{\tilde{Y}(1), \tilde{Y}(0), \tilde{Y}(1)-\tilde{Y}(0), \tilde{{\varepsilon} } (1),$ $\tilde{{\varepsilon} } (0), \tilde{{\varepsilon} } (1)-\tilde{{\varepsilon} } (0)\})$ 是对应群组变量的方差，例如，$S^2_{\tilde{Y}(z)} = (m-1)^{-1} \sum_{i=1}^{m} \big\{\tilde{Y}_{i\cdot}(z)-m^{-1}\sum_{j=1}^{m}\tilde{Y}_{j\cdot}(z)\big\}^2$。令$\tilde{\omega}_i =n_{(i)} m/n$和$\tilde{\omega} = (\tilde{\omega}_1,\ldots, \tilde{\omega}_m)^\top$。

在群组随机化试验中，可以观测到个体水平或群组水平的协变量。令 $\bs{x}_{ij}=(x_{ij1},\ldots,x_{ijp})^{\top}$ 表示个体 $(i,j)$ 的协变量，$\bs{c}_i=(c_{i1},\ldots,c_{ip^\prime})^{\top}$ 表示群组 $i$ 的协变量。群组水平和个体水平协变量可以互相转化。
一方面，可以通过群组内的个体水平协变量的某些描述性统计量来构造群组水平的协变量。例如，可以定义 $\bs{c}_{i} = \sum_{j=1}^{n_{(i)}} \bs{x}_{ij}/n_{(i)}$。
另一方面，若仅能在重随机化前观测到群组水平的协变量 $\bs{c}_i$，则可以对所有的 $j=1,\ldots,n_{(i)}$，定义个体水平的协变量为 $\bs{x}_{ij}=\bs{c}_i$。因此我们考虑先将协变量统一转化为群组水平变量或者个体水平变量后再研究基于群组水平变量的重随机化和基于个体水平变量的重随机化。基于群组水平协变量的重随机化使用群组水平变量的均值差：$\htcht   =m_1^{-1} \sum_{i=1}^m  Z_i \bs{c}_i - m_0^{-1}\sum_{i=1}^m  (1-Z_i)\bs{c}_i$。
基于个体水平协变量的重随机化使用个体水平协变量的均值差：$\htxhaj   = n_1^{-1} \sum_{i=1}^m  Z_i\sum_{j=1}^{n_{(i)}}\bs{x}_{ij} -n_0^{-1}  \sum_{i=1}^m  (1-Z_i)\sum_{j=1}^{n_{(i)}}\bs{x}_{ij}$。

Lu等人 \citep{Lu2023} 研究了基于$\htcht$和$\htxhaj$的马氏距离重随机化以及更一般的基于欧式距离的重随机化，这里我们简要介绍如下马氏距离重随机化：
\begin{equation*}
 \mc  = \{  \htcht ^{\top} \cov (\htcht) ^{-1}\ \htcht  \leq a \} ,\quad \mx =  \{  \htxhaj ^{\top} \cov (\htxhaj) ^{-1}\ \htxhaj \leq a \}.
\end{equation*}

我们考虑 $\htht$和$\mc$的组合以及$\hthaj$和$\mx$的组合，因为前者都基于群组水平的变量，后者都基于个体水平的变量。我们有如下关于重随机化下点估计的渐近分布的性质。记$R^2_{\bs{c}}$和$R^2_{\bs{x}}\in [0,1]$ 分别为 $\htht$和 $\htcht$、$\hthaj$和$\htxhaj$之间的相关系数的平方。令$L_{p,a}\sim D_1\mid \bs{D}^\top\bs{D}\leq a$，其中 $\bs{D}=(D_{1}, \ldots, D_{p})^{\top}$ 是 $p$ 维标准正态随机向量，并且令 $\epsilon$ 为与 $L_{p,a}$ 独立的标准正态随机变量。令$\tilde{\bs{x}}_{i\cdot} = \sum_{j=1}^{n_{(i)}} \bs{x}_{ij} m/n$。

\begin{theorem}[\citep{Lu2023}]
\label{thm:asymp2}
    在群组随机化试验中若$\lim_{m\rightarrow \infty} m_1/m = C\in (0,1)$，有限总体方差和协方差 $S^2_{\tilde{Y}(z)}$ 、$S^2_{\tilde{\varepsilon }(z)}$ 、$\bs{S}_{\bs{c},\tilde{Y}(z)}$、$\bs{S}_{\tilde{\bs{x}},\tilde{\varepsilon } (z)} $ $(z=0,1)$、  ${S}_{\tilde{Y}(0),\tilde{Y}(1)} $、 ${S}_{\tilde{\varepsilon }(0),\tilde{\varepsilon }(1)}$、 $\bs{S}^2_{\bs{c}}$、$\bs{S}^2_{\tilde{\bs{x}}}$、$S^2_{\tilde{\omega}}$、  $\bs{S}_{\tilde{\bs{x}},\tilde{\omega}}$ 有有限的极限值，$\bs{S}^2_{\bs{c}}$、 $\bs{S}^2_{\tilde{\bs{x}}}$ 的极限是非奇异的，$m\sigma^2_{\textnormal{haj}}$和$m\sigma^2_{\textnormal{ht}}$ 的极限为正，且 
    $$m^{-1} \max_{z\in\{0,1\}}\max _{1 \leq i \leq m}\Big[ \{\tilde{Y}_{i\cdot}(z)-\bar{Y}(z) \}^2,  \big\{\tdei(z)\big\}^2 ,  \|{\bs{c}}_{i} \|_{\infty}^{2}  , \|\tilde{\bs{x}}_{i\cdot}\|_{\infty}^2 \Big] \rightarrow 0,$$
则有
    \begin{align*}
    m^{1/2}(\hthaj  -\tau) \mid \mx &\mathrel{\ \dot{\sim}\ }  (m\sigma^2_{\textnormal{haj}})^{1/2}\{(1-R_{\bs{ x}}^{2})^{1/2}  \epsilon+R_{\bs{x}}  L_{p, a}\},\\
    m^{1/2}(\htht  -\tau) \mid \mc  &\mathrel{\ \dot{\sim}\ }  (m\sigma^2_{\textnormal{ht}})^{1/2}\{(1-R_{\bs{ c}}^{2})^{1/2}  \epsilon+R_{\bs{c}} L_{p^\prime, a}\}.
\end{align*}
若$\bs{c}_i = (n_i, \tilde{\bs{x}}_{i\cdot} )$，则有 $(m\sigma^2_{\textnormal{ht}})(1-R_{\bs{c}}^{2}) \leq (m\sigma^2_{\textnormal{haj}})(1-R_{\bs{ x}}^{2}) $。
\end{theorem}
当重随机化的阈值较小时，$m^{1/2}(\hthaj  -\tau)$和$m^{1/2}(\htht  -\tau)$的渐近方差近似为
$(m\sigma^2_{\textnormal{ht}})(1-R_{\bs{c}}^{2})$和$(m\sigma^2_{\textnormal{haj}})(1-R_{\bs{ x}}^{2})$。因此，定理\ref{thm:asymp2}表明，当群组水平协变量包括群组大小和缩放的个体水平协变量时，基于群组水平协变量的重随机化方法优于基于个体水平协变量的重随机化方法。

Lu等人 \citep{Lu2023}还比较了基于马氏距离的重随机化方法和两种常用的使用协变量重要程度先验信息的重随机化方法：基于加权欧式距离和基于分层马氏距离的重随机化方法。该论文证明了在协变量正交化后且使用最优的权重的情况下,基于加权欧式距离的重随机化方法是最优的。此外，群组重随机化也可以和回归调整结合起来使用。Lu等人 \citep{Lu2023}证明了在大样本下，Lin的带有交互相的回归仍然适用。不过一些渐近结果与个体水平处理分配下的重随机化结果不同。具体而言，当采用群组重随机化并使用个体水平的协变量进行回归调整时，其渐近分布将不再服从正态分布，二是正态分布与截断正态分布的卷积。

Lu等人 \citep{Lu2023}建立的渐近理论要求有大量的群组，而这在许多群组随机化试验中可能并不现实。当群组个数$m$较小时，一种替代方法是使用混合效应模型，即对数据生成过程施加参数化假设。另一种替代方法是使用 Fisher 精确检验，该方法在强零假设下可以给出有限样本精确的$p$-值。有关更多细节，可参见Zhao和Ding\citep{zhao2021covariate}。需要特别注意的是，Fisher 精确检验必须遵循与群组重随机化相同的处理分配规则。当群组个数为中等大小时，可以使用更加稳健的ToM回归调整估计量\citep{Lu2024}。

\subsection{裂区试验和裂区重随机化试验}

裂区试验（split-plot experiments）是一种特殊的因子试验，被广泛应用于农业、工业、社会科学以及生物医学领域，用于处理难以变更的因子。该试验作为一种研究多个干预因子的析因设计，其试验个体（子区）嵌套于更高层级的分组（主区）之中，其干预操作分两阶段进行。第一阶段：通过主区层级的群组随机化分配部分因子（称为主区因子）；第二阶段：在子区层级通过分层随机化分配剩余因子（称为子区因子）。

根据设计，同一主区内的所有子区都会接受相同水平的主区因子，这种结构便于处理不易变更的试验因子，同时避免了主区内的干扰效应。但另一方面，主区内子区之间的依赖结构也为统计推断带来了挑战\citep{wu2011experiments}。 Kempthorne\cite{kempthorne1952design} 开启了关于裂区试验下基于设计的推断讨论，该研究假设个体处理效应为常数。Zhao等人 \citep{zhao2018randomization} 放宽了该假设，在子区处理组个体数量在所有主区中保持一致的条件下，建立了基于样本均值的估计量的有限样本精确推断理论。Mukerjee 和 Dasgupta \citep{mukerjee2019causal} 将讨论拓展到主区间处理组个体数量不一致的裂区试验，并对Horvitz--Thompson估计量给出了有限样本精确推断理论。Zhao和Ding \citep{zhao2021reconciling} 进一步将理论扩展到Hajek估计量，并建立了Horvitz--Thompson与Hajek估计量的相合性与渐近正态性。

在裂区试验中，试验者常常会在主区与子区层级收集基线协变量信息。例如，以学生为子区个体、以班级为主区的裂区试验中，班级规模和教师经验属于主区协变量，而学生的年龄和性别属于子区协变量。上文讨论的重随机化有利于解决协变量不平衡的问题。但是，已有关于重随机化的文献主要聚焦于单一层级（个体层或组层）的处理分配，对于裂区试验的理论探讨仍留有空白。为填补这一空白，Shi等人 \citep{Shi2025}提出了两种适用于裂区试验的重随机化策略，建立了Horvitz--Thompson和Hajek估计量的渐近理论，证明了重随机化可以提高平均处理效应的估计和推断效率。Shi等人 \citep{Shi2025}还研究了Zhao和Ding \citep{zhao2021reconciling}  所推荐的回归调整方法，
证明了其在仅使用主区协变量的裂区重随机化中也能带来效率提升。

类似于群组重随机化，裂区重随机化下回归调整估计量的渐近分布并非正态，而是正态分布与截断正态分布的卷积。此外，Zhao和Ding \citep{zhao2021reconciling}  还发现若使用异质的子区协变量进行回归调整，可能会降低精度。为了克服回归调整的这一局限性，Shi等人 \citep{Shi2025}提出了一种基于条件偏差修正的协变量调整策略。该方法依据协变量平衡性对估计量进行渐近偏差修正，所得估计量在裂区重随机化下具有相合性与渐近正态性，并且有效率提升的理论保证。这些理论结果都是基于设计的，允许线性模型误设。

\subsection{缺失数据问题}


在实际应用中，协变量和结果变量都可能存在缺失值，这给回归调整的实施带来了挑战。针对协变量缺失问题，已有多种方法被提出，包括最大似然估计、贝叶斯方法和多重插补等\citep{rubin2004multiple}。但这些方法的有效性强依赖于其假设的有效性。如果假设不成立，这些方法可能会造成精度损失。

在完全随机化和基于设计的框架下，Zhao和Ding \citep{Zhao:2022}分析了一系列缺失值处理方法在Fisher回归和Lin回归下的性质，证明了在Lin回归下，这些缺失值处理方法相比于不进行协变量调整都有精度的提升，并给出了这些缺失值处理方法在理论上的渐近方差的大小关系。
进一步地，Zhao等人 \citep{Zhao-Ding-biometrika}探究了在协变量和结果变量同时存在缺失时的处理方法。该论文证明了在此情况下回归调整和逆概率加权并不等价，前者并不确保精度提升，而后者则可以确保加入更多协变量一定不会降低估计精度。
为了解决Lin回归参数过多、不够稳健的问题，Fu等人 \citep{Fu:2025}证明了ToM回归也可以用于解决协变量缺失的问题。



\section{高维回归调整}
\setcounter{section}{5}
\setcounter{theorem}{0}
\setcounter{assumption}{0}
\setcounter{definition}{0}

本章考虑当协变量是高维时，进行协变量调整的方法和理论。

\subsection{Lasso回归调整}

在相关性分析中，高维变量选择方法（如 Lasso \citep{Tibshirani:1996}）在参数估计、变量选择、响应预测和统计推断等方面发挥着重要作用。在因果推断领域，这类高维统计与机器学习方法同样具有重要影响力。例如，在完全随机化与分层随机化试验中，Lasso 回归调整为处理高维协变量提供了一条有效途径。该方法在有限总体框架下不仅保留了随机化推断的严谨性，而且通过稀疏正则化避免了传统回归在高维场景中的过拟合问题，从而在理论与实践中都展现出重要价值。

Bloniarz等人 \citep{Bloniarz2016} 在完全随机化试验下提出了基于 Lasso 的回归调整方法。该方法首先在处理组和对照组内分别拟合 Lasso 回归  
$$
\hat {\bs\beta}_{z,\text{lasso}} = \arg\min_{\bs{\beta} \in \mathbb{R}^p} 
\left\{ \frac{1}{2n_z}\sum_{i:Z_i=z} \big(Y_i - \bs{x}_i^\top \bs{\beta} \big)^2 + \lambda \|\bs{\beta}\|_1 \right\}, \quad z \in \{0,1\},
$$
其中，$||\cdot||_1$是向量的$\ell_1$范数，$\lambda\geq 0$是正则化参数，其具体数值可以通过交叉验证来选取。然后，计算Lasso调整后的平均处理效应估计量： 
$$
\hat\tau_{\text{lasso}} = (\bar Y_1 - \bar{\bs{x}}_1^\top \hat{\bs{\beta}}_{1,\text{lasso}}) - (\bar Y_0 - \bar{\bs{x}}_0^\top \hat{\bs{\beta}}_{0,\text{lasso}}).
$$
直观上，这一做法利用 Lasso 从高维协变量中筛选出与结果变量最相关的变量，并用残差替代原始结果变量，以减少由协变量不平衡引起的噪声。注意到，此处的Lasso回归只对协变量系数$\bs{\beta}$施加了$l_1$惩罚，而没有对处理效应施加惩罚，因此仍然可以相合地估计处理效应并进行渐近推断。

在理论层面，Bloniarz等人 \citep{Bloniarz2016} 在基于设计的框架下证明了当潜在结果和协变量满足一定的矩条件，两者之间的线性投影系数满足稀疏性条件且正则化参数选择得当时，Lasso 调整后的处理效应估计是相合的，并且其极限分布为正态分布。更重要的是，Lasso 调整后的估计量的渐近方差小于等于无调整估计量的渐近方差，从而具有精度优势。此外，他们提出了基于组内残差平方和的方差估计方法，可以在有限样本下给出保守的方差估计，并由此构造渐近有效的 Wald 型置信区间。这些结果表明，在完全随机化试验下，Lasso 回归调整不仅解决了高维场景下 OLS 不可行的问题，而且在理论上保证了推断的有效性和效率提升。

针对更复杂的分层随机化与重随机化，Zhu等人 \citep{zhu2025design} 将 Bloniarz等人提出的 Lasso 回归调整进行了系统性推广，并在基于设计的框架下建立了严格的大样本理论。Zhu等人 \citep{zhu2025design}提出的方法的核心思想是：将分层加权的均值差$\hat\tau_{\text{unadj}}$投影到分层加权的协变量均值差$\hat{\bs{\tau}}_{\bs{x}}$上（定义参见\eqref{eq:sr-unadj}和\eqref{eq:sr-x}），得到理想投影估计量（oracle projection estimator）$\check{\tau}_{\text{proj}}=\hat\tau_{\text{unadj}}-\hat{\bs{\tau}}_{\bs{x}}^\top\bs{\gamma}_{\text{proj}}$。由于$E(\hat{\bs{\tau}}_{\bs{x}})=\bs{0}$并且投影以最小方差为目标, 这一构造在保持估计量相合性的同时，可以实现在一类估计量$\{\hat\tau_{\text{unadj}}-\hat{\bs{\tau}}_{\bs{x}}^\top\bs{\gamma}$，$\bs{\gamma}\in\mathbb{R}^p\}$下的最小方差。通过使用Lasso回归对投影向量$\bs{\gamma}_{\text{proj}}$进行估计，我们可以在不依赖潜在结果模型正确性的前提下，从噪声中剥离出与潜在结果最相关的协变量成分，从而降低方差。

记$\hat\tau_{\text{stra-lasso}}=\hat\tau_{\text{unadj}}-\hat{\bs{\tau}}_{\bs{x}}^\top \hat{\bs{\gamma}}_{\text{proj}}$为Lasso调整后的估计量，$\sigma^2_{\text{stra-lasso}}$为其渐近方差。
Zhu等人 \citep{zhu2025design} 在允许层间处理比例异质、层大小异质、甚至“细分层”（某些层中仅有一个处理或对照个体）存在的宽松条件下，证明了
$
\sqrt{n}(\hat\tau_{\text{stra-lasso}}-\tau)\ \xrightarrow{d}\ \mathcal N(0,\ \sigma^2_{\text{stra-lasso}})
$，
并且给出了一个 Neyman 型的保守方差估计 $\hat\sigma^2_{\text{stra-lasso}}$，其在一定正则性条件下渐近不小于 $\sigma^2_{\text{stra-lasso}}$，从而可用于构造渐近覆盖率不低于指定水平的置信区间与检验。Lasso 调整在渐近方差上不超过未调整的分层均值差估计量，且在协变量对结果变量有解释力（即投影系数稀疏且非零）的情形下严格降低渐近方差。此外，在设计阶段若进一步结合分层重随机化（例如以加权协变量均值的马氏距离作为准则，设其接受集为 $\mathcal M_a$；参见 Wang等人\citep{Wang2023}），Zhu等人 \citep{zhu2025design} 证明了条件于接受集 $\mathcal M_a$ 的随机化分布下有
$
\{\sqrt{n}(\hat\tau_{\text{stra-lasso}}-\tau)\ |\ \mathcal M_a\}\ \xrightarrow{d}\ \mathcal N(0,\ \sigma^2_{\text{stra-lasso}})
$，且由设计强化带来的协变量平衡会在有限样本下提升估计和推断的效率。与完全随机化情形相比，这一结果的适用面更广：它既覆盖异质处理比例与异质分层规模，也允许极端细分层结构而无需额外的模型化假设。综合而言，分层随机化与分层重随机化提供了“先平衡、后调整”的双重增益；在分析阶段实施 Lasso 回归调整，则在有限总体推断下以投影最优的方式系统降低方差，实现对传统分层均值差估计的稳健改进。


这些研究说明，Lasso 回归调整在高维协变量背景下为完全随机化和分层随机化试验都具有坚实的理论基础。无论是简单的均匀随机分配，还是复杂的分层与重随机化设计，基于稀疏正则化的调整方法都能在有限总体框架下实现有效推断，并在效率上优于未调整的估计量，从而为存在高维协变量的随机化试验分析提供了统一而灵活的工具。

\subsection{去偏回归调整}

回归调整估计量（如Lin回归、ToM回归）在协变量维度随样本量趋向无穷大时，会产生显著偏差。具体而言，Li和Lei \citep{Lei2020}证明：受偏差项影响，若要保证Lin回归调整估计量的相合性，需满足协变量维度条件 $p = o(n/\log n)$；若进一步要求估计量的渐近正态性，则需更强的条件 $p = o(n^{1/2})$。但在实际随机化试验中，研究者往往会测量大量协变量，导致 $p$ 与 $n$ 可比甚至 $p \gg n$。偏差是重要的统计学指标，在很多实际应用中具有重要意义。因此，如何解决回归调整的偏差问题以适配高维协变量场景，已成为该领域的重要研究方向。

第5.1节介绍的Lasso回归调整方法虽能处理 $p\gg n$ 的高维情形，但其有效性依赖于难以验证的稀疏性假设——即潜在结果仅由协变量的稀疏线性组合决定。不依赖强假设的偏差校正方法因此受到研究者关注。这类方法通常通过减掉Lin回归调整估计量的偏差估计，实现偏差缩减。其中，Lei和Ding \citep{Lei2020}提出的去偏估计量可在 $p = o(n^{2/3})$ 时保证渐近正态性；Chang等人\citep{chang2024exact}提出的方法则将适用范围扩展至 $p = o(n)$；而Lu等人\citep{lu2023debiased}提出的去偏回归调整估计量，更可处理 $p < n-1$ 的高维场景。

\subsection{模型平均方法}

协变量调整并不局限于使用线性回归模型，还可以使用其他的机器学习模型，比如决策树、随机森林、核岭回归、神经网络与深度学习等。并且由于基于设计的因果推断不需要模型的正确指定，因此可以将多个模型的优势结合起来使用，并使用模型平均等方法选取最优的组合系数。

模型平均是一类统计方法，用于在存在多个候选模型时，通过加权平均来综合它们的预测或估计结果，从而降低模型选择不确定性对推断的影响\citep{Buckland1997,Hansen2007,RollingYang2014}。与单一模型选择不同，模型平均不需要完全依赖某个``最佳"模型，而是通过模型集合的综合来提高估计的稳定性和效率。在抽样调查和因果推断中，尤其是平均处理效应的估计中，模型平均可以用来结合不同的调整方法（如回归调整、分层、倾向得分加权等），在保持估计相合性的同时提高精度。Breidt 和 Opsomer \citep{Breidt2017}综述了在抽样调查中，如何利用现代预测方法（包括集成与模型平均）作为辅助，提升估计效率；Aronow 和 Middleton\citep{Aronow2013}在基于设计的框架下，指出任意无偏的平均处理效应估计量的凸组合仍然无偏，这为随机化试验中使用模型平均方法提供了理论保证。

\section{不依从问题}
\setcounter{section}{6}
\setcounter{theorem}{0}
\setcounter{assumption}{0}
\setcounter{definition}{0}


正如上文所述，随机化试验广泛应用于社会科学、医学研究及诸多其他领域，以揭示因果关系。但在实践中，部分试验个体可能不遵守所分配的处理方案。例如，某人被随机分配服用新药，但他/她仍有权拒绝该药而服用安慰剂。不依从问题在各类随机试验中普遍存在，如临床试验、社会试验、政策评估和教育研究。由于试验个体仅被分配到其中一个处理组，我们无法观察到他们在另一个处理组下实际接受处理的状态。因此，我们无法知晓谁是依从者，也难以解释所估计的处理效应。

潜在结果框架的一个成功应用是阐明了当存在不依从问题时识别处理效应所需的基本假设\citep{Imbens:1994, Angrist:1995, Angrist:1996, Imbens:1997}。Angrist等人\citep{Angrist:1996}提出了一组识别条件，在该条件下，依从者平均处理效应（Complier Average Treatment Effect; CATE）可以识别为两个意向性治疗效应（intention-to-treat effect; ITT effect）的比值：处理分配对潜在结果的平均因果效应与处理分配对实际处理接受状态的平均因果效应。因此，可以使用插入法，通常称为Wald估计量，或者两阶段最小二乘法\citep{Angrist:2008}估计依从者平均处理效应。
在超总体框架下，Imbens 和 Angrist \citep{Imbens:1994}证明了Wald估计量具有相合性和渐近正态性。
Ren \citep{Ren2024cate}在有限总体和随机化推断的框架下，为Wald估计量的渐近正态性提供了严格的证明。此外，Wald估计量并未纳入协变量信息。为解决此问题，计量经济学中最流行的方法是使用带有协变量的两阶段最小二乘法\citep{Angrist:2008}。此外，在完全参数化模型设定下，最大似然法和贝叶斯推断方法也被用于估计CATE \citep{Imbens:1997}。然而，如果工作模型错误指定，这些方法将缺乏有效性。半参数方法也被用于估计CATE\citep{Abadie:2003, Wang2023}。但这些研究都假设观测数据是来自某个超总体的独立同分布的样本。Ren \citep{Ren2024cate}则考虑了有限总体和基于随机化推断的框架，通过纳入协变量信息提出了三种模型辅助的CATE估计量，研究了它们的渐近性质，并将其效率与Wald估计量进行了比较。该论文的理论分析允许工作模型任意错误指定。此外，该论文还提供了保守的方差估计量以构建CATE的渐近保守置信区间。

\section{网络干扰}
\setcounter{section}{7}
\setcounter{theorem}{0}
\setcounter{assumption}{0}
\setcounter{definition}{0}

\subsection{概述}

除了不依从问题外，分析随机化试验的另一个难点在于试验个体之间可能存在相互影响\citep{holtz2020interdependence, jiang2023auto}。例如，在一项关于金融知识普及的试验中，研究者随机向村庄部分家庭提供更为详尽的讲座，以考察其对保险购买意愿的影响。但由于社交交往，接受详细讲座的家庭往往会进一步影响接受普通讲座家庭的购买决策 \citep{cai2015social}。一般而言，当个体的潜在结果受到他人处理分配影响时，即被称为存在干扰（interference）。此类干扰通常源于社交网络，如家庭、社区或朋友关系 \citep{aral2017exercise, jiang2023auto}。在干扰存在的情况下，稳定个体处理值假设不再成立，这使得传统的均值差或者Horvitz–Thompson估计量产生偏差。因此，一个核心问题是如何在干扰环境下有效估计感兴趣的因果效应，并建立相应的推断方法。

处理干扰问题的常见策略之一是使用群组随机化试验。在估计全局平均处理效应（global average treatment effect; GATE），即所有个体均接受处理与所有个体均不接受处理两种情形下平均结果的差异时，如果群组划分得当，该方法能够消除由干扰带来的偏差。然而，现实中的网络结构通常极为复杂，高质量的群组划分难以实现。例如，Saveski 等人\citep{saveski2017detecting} 提出的最优群组划分策略仅覆盖约$35.59\%$的网络边，说明即便采用群组随机化，偏差仍难以彻底消除。另一方面，尽管群组随机化在社会科学研究中较为常见，但在eBay、Facebook和Twitter等平台的实践中，更为普遍的仍是操作简便的伯努利试验。此外，群组随机化还存在一个重要局限：它无法区分并识别直接平均处理效应（direct average treatment effect; DATE）和间接平均处理效应（indirect average treatment effect; IATE）。然而，在诸多应用中，这两类效应至关重要，例如评估某地区政策如何通过网络扩散而影响自身以及其他地区个体的行为 \citep{holtz2020interdependence}。

近年来，关于网络干扰下因果推断的研究逐渐增多\citep{hudgens2008toward, aronow2017estimating, leung2022causal}。Chin \citep{chin2019regression} 提出了基于普通最小二乘回归的GATE估计方法，但其工作局限于超总体框架，难以直接应用于有限总体的基于设计的推断。另一类研究依赖于暴露映射（exposure mapping），假设个体潜在结果仅通过某个已知的低维映射(例如个体自己的处理状态以及该个体的网络邻居中接受处理的比例)依赖于自身和他人的处理分配\citep{aronow2017estimating, harshaw2023design}。这类方法解决了处理向量取值空间过大的问题，但其有效性高度依赖于暴露映射的正确指定，并且往往对网络干扰的密集程度有所限制。为了解决这一问题，Leung \citep{leung2022causal} 提出了近邻干扰假设，放宽了对暴露映射正确指定的要求，仅假设干扰效应随网络距离增大而衰减。然而，该框架主要适用于稀疏网络，在稠密网络中表现不佳。此外，还有一些研究尝试通过新的试验设计来识别和估计处理效应。例如，饱和设计（saturation design，Baird等人\citep{baird2018optimal}，群组随机化的扩展）、二部设计（bipartite design，Harshaw等人\citep{harshaw2023design}）、群组化网络试验（clustered network experiment，Chattopadhyay等人\citep{chattopadhyay2023design}）、以及多重随机化设计（multiple randomization design，Masoero等人\citep{masoero2024multiple}）。不过，这些设计通常依赖于对干扰结构的特殊假设，例如部分干扰（partial interference）或局部干扰（local interference）假设，因此无法适用于一般的网络干扰情形。此外，这些研究对处理效应的识别方法也与试验设计高度绑定，缺乏在 DATE、IATE 与 GATE 的估计上统一的方法论框架，也难以直接应用于伯努利试验。

\subsection{Fisher精确检验}

在不存在干扰的情形下，Fisher 精确检验能够在有限样本下精确有效，因为强零假设唯一确定所有潜在结果 \citep{Fisher1935}。然而，在网络干扰普遍存在的实际应用中，经典 Fisher 精确检验 不再可行：通常感兴趣的原假设无法填补所有潜在结果。围绕这一难题，近年来学界主要沿着两条思路对Fisher 精确检验 进行了扩展：一类是条件随机化检验，另一类是基于填补的随机化检验。

条件随机化检验的核心思想是将注意力限制在某些子样本和子分配空间中，以恢复强零假设。具体做法是选取一部分“焦点个体”，并同时规定一组“焦点分配”，使得在该条件化的试验空间中，感兴趣的原假设可以填补所有计算检验统计量需要的潜在结果，从而进行后续的随机化检验 \citep{aronow2012general,athey2018exact,basse2019randomization,puelz2022graph,zhang2025multiple}。这一方法在理论上保证了因果推断的有限样本有效性，然而，条件随机化检验必须在焦点个体与焦点分配上做出限制，部分样本信息被舍弃，导致检验功效不足。后续研究提出了若干改进方法。例如Zhong\cite{zhong2024unconditional} 提出了部分零假设随机化检验，通过比较观测分配与所有可能分配下统计量的差异，避免了对焦点分配的条件化，从而一定程度上提升了功效。然而，当焦点个体比例很低时，部分零假设随机化检验依旧存在明显的功效不足，这也表明 条件随机化检验相关方法在复杂网络环境下仍面临较大挑战。

另一条发展路径借鉴了 Rubin 提出的多重填补思想 \citep{rubin1977formalizing,rubin2004multiple}。多重填补已广泛应用于因果推断中的各种“未知”处理，如填补依从性状态以增强非依从试验下的 Fisher 精确检验的功效 \citep{rubin1998more}，在部分事后亚组分析中填补潜在结果 \citep{lee2015valid}，处理因子效应不完全可识别的问题 \citep{espinosa2016bayesian}，应对缺失结局 \citep{ivanova2022randomization}，弥补公共卫生试验中疫情冲击带来的数据缺口 \citep{uschner2023using}等。基于这一思路，Han等人\citep{han2024imputation} 提出了适用于网络干扰情境的基于填补的随机化检验方法。其基本做法是：在原假设下为无法观测的潜在结果构造合适的填补分布，并据此多次生成完整数据集；在每一次完整数据上执行Fisher 精确检验，得到一系列 $p$ 值；最后将这些 $p$ 值进行平均，作为最终的检验 $p$ 值。与条件随机化检验不同，基于填补的随机化检验保留了所有个体和所有可能的处理分配，因而大大提升了检验功效。理论上，在填补分布合理指定的条件下，基于填补的随机化检验能够在渐近意义下将第一类错误率控制在不超过显著性水平两倍的范围内。数值结果进一步显示，该方法在实际有限样本中通常能维持接近指定水平的错误率，并在功效上显著优于条件随机化检验和部分零假设随机化检验。

由此可见，条件随机化检验与基于填补的随机化检验分别代表了在干扰环境下扩展Fisher 精确检验的两条研究路线：前者强调通过条件化保持有限样本的精确性；后者则通过填补恢复强零假设，从而在保证错误率可控的同时显著提升功效。这些进展表明，Fisher 精确检验在网络干扰下仍然可以通过合理的扩展发挥作用，并且在理论与实践中都展现出越来越重要的价值。

\subsection{Neyman渐近推断和回归调整}
当存在溢出效应时，数据会呈现出极强的相依结构，这种结构贯穿于网络、协变量、处理变量及潜在结果的生成过程中，导致数据随机性的建模与刻画面临显著挑战。
在溢出效应场景中，个体的潜在结果会依赖于试验中所有个体的处理状态。以包含$n$个个体（记个体集合为$\mathcal{N}_n = \{1,2,\ldots,n\}$）的二元处理随机化试验为例：
若不施加任何限制，每个个体的潜在结果会受所有个体处理组合的影响，因此单个个体将存在$2^n$种潜在结果。
对于第$i$个个体（$i\in\mathcal{N}_n$），其潜在结果可表示为$Y_i(\bs{z})$，其中$\bs{z}\in \{0,1\}^n$是所有个体的潜在处理向量（$\bs{z}$中元素为1表示对应个体接受处理，为0则表示接受对照）。然而，在单次随机化试验中，我们仅能观察到$2^n$种潜在结果中的一种：当试验实现的处理向量为$\bs{Z} = (Z_1,Z_2,\ldots,Z_n)$时，实际可观测的结果仅为$Y_i = Y_i(\bs{Z})$（即与$\bs{Z}$对应的潜在结果）。这种“仅能观测单次实现结果”的特性，为溢出效应下的因果分析带来了核心挑战。

为解决上述挑战，Aronow 和 Samii \citep{aronow2017estimating}提出了暴露映射假设，其核心思想是：$n$维的高维处理向量，仅通过一个低维映射影响潜在结果，从而将高维问题转化为低维问题。具体而言，假设个体之间通过某个网络而互相影响。记该网络的邻接矩阵为$\bs{E} = (E_{ij})_{i,j\in\mathcal{N}_n}\in \{0,1\}^{n\times n}$，其中$E_{ii}=0$且$E_{ij}=1$代表第$j$个个体的处理可能影响第$i$个个体的潜在结果。令$\mathcal{E}_n$为所有可能的邻接矩阵构成的集合。暴露映射的严格定义为：  
$\bs{D}:\mathcal{N}_n\times \{0,1\}^n\times \mathcal{E}_n\rightarrow \mathcal{D}$，  
其中$\mathcal{D} \subseteq \mathbb{R}^{d_{\bs{D}}}$是维度为$d_{\bs{D}} \in \mathbb{N}$的有限集合。基于此，Aronow 和 Samii \citep{aronow2017estimating}提出暴露映射假设：

\begin{assumption}\label{a:exposure-mapping}
对任意个体$i\in \mathcal{N}_n$及任意两个处理向量$\bs{z}, \bs{z}^\prime \in \{0,1\}^n$，若二者满足$\bs{D}(i,\bs{z},\bs{E}) = \bs{D}(i,\bs{z}^\prime,\bs{E})$，则$Y_i(\bs{z}) = Y_i(\bs{z}^\prime)$。
\end{assumption}

该假设意味着，潜在结果仅由暴露映射的结果决定，而非高维处理向量本身。因此，可将潜在结果简化表示为$\tilde{Y}_i(\bs{d})$（$\bs{d}\in \mathcal{D}$），其中$\bs{d}$是暴露映射的具体结果。常见的一种暴露映射为$\bs{D}(i,\bs{z},\bs{E}) = (z_i,{\sum_{j=1}^n E_{ij}z_j}/{\sum_{j=1}^n E_{ij}})$，即同时包含“个体自身的处理状态”与“其邻居的处理比例”两个低维信息。基于暴露映射假设，Aronow 和 Samii \citep{aronow2017estimating}定义了待估因果效应$\tau$，即两种暴露状态下潜在结果的均值差：  
$$\tau = \mu(\bs{d}) - \mu(\bs{d}^\prime),\quad \mu(\bs{d}) = \frac{1}{n}\sum_{i=1}^n \tilde{Y}_i(\bs{d}),$$  
其中$\bs{d}$与$\bs{d}^\prime$是两种不同的暴露状态，$\mu(\bs{d})$是所有个体在暴露状态$\bs{d}$下的潜在结果均值。Aronow 和 Samii \citep{aronow2017estimating}建议使用Horvitz--Thompson（HT）估计量来估计$\tau$。令$\bs{D}_i = \bs{D}(i,\bs{Z},\bs{E})$为个体$i$的暴露映射实现，$\pi_i(\bs{d}) = \mathbb{P}(\bs{D}_i = \bs{d})$，则$\tau$有如下无偏估计：
\[
\hat{\tau} = \hat{\mu}(\bs{d}) - \hat{\mu}(\bs{d}^\prime),\quad \hat{\mu}(\bs{d}) = \frac{1}{n} \sum_{i=1}^n\frac{{Y}_i I(\bs{D}_i = \bs{d})}{\pi_i(\bs{d})}.
\]

Aronow 和 Samii \citep{aronow2017estimating}提出的方法的核心局限性在于：暴露映射假设是一个理想条件，其正确性在实际研究中难以验证，而待估量的定义、识别与推断均高度依赖该假设。为了解决这个难题，Leung \citep{leung2022causal}提出了近邻干扰假设，在保留暴露映射概念的同时，不再强依赖暴露映射假设。近邻干扰假设的核心思想是：个体受到其他个体影响而产生的溢出效应会随着二者在网络中距离的增加而逐渐衰减（即距离越远，溢出效应越弱，直至可忽略）。

由于不再假设暴露映射假设成立，潜在结果无法简化为$\tilde{Y}_i(\bs{d})$，因此需要重新定义待估因果效应$\tau$：  
$$\tau = \mu(\bs{d}) - \mu(\bs{d}^\prime),\quad \mu(\bs{d}) = \sum_{\bs{z}\in \{0,1\}^n} \mathbb{P}(\bs{Z} = \bs{z}\mid \bs{D}_i=\bs{d})  Y_i(\bs{z}).$$  
可以验证：Aronow 和 Samii \citep{aronow2017estimating}提出的$\hat{\mu}(\bs{d})$仍是上述新定义下$\mu(\bs{d})$的无偏估计量，因此$\hat{\tau} = \hat{\mu}(\bs{d}) - \hat{\mu}(\bs{d}^\prime)$依然适用，且可进一步改写为$\hat{\tau} = n^{-1} \sum_{i=1}^n U_i$，其中$U_i = {Y_i  I(\bs{D}_i = \bs{d})}/{\pi_i(\bs{d})} - {Y_i I(\bs{D}_i = \bs{d}^\prime)}/{\pi_i(\bs{d}^\prime)}$。为解决溢出效应导致的自相关问题，Leung \citep{leung2022causal}提出了异方差和自相关一致（Heteroskedasticity and Autocorrelation Consistent; HAC）方差估计量：  
$$\hat{\sigma}^2 = \frac{1}{n^2}\sum_{i=1}^n \sum_{j=1}^n (U_i - \hat{\tau})(U_j - \hat{\tau}) \cdot I(\ell_{\bs{E}}(i,j) \leq b_n).$$  
其中，$\ell_{\bs{E}}(i,j)$是个体$i$与$j$在网络$\bs{E}$中的距离；$b_n$是依赖于样本量$n$的带宽参数，满足$b_n \to \infty$。Leung \citep{leung2022causal}在特定条件下证明了估计量$\hat{\tau}$具有渐近正态性（即当$n$足够大时，$\hat{\tau}$的分布近似服从正态分布），且方差估计量$\hat{\sigma}^2$在一定条件下具有保守性（即在渐近的意义下，估计的方差不会低估真实方差，保证了统计推断的渐近有效性）。

在存在干扰的情况下，当观测到个体协变量$\bs{x}_i$时，仍然有可能通过回归调整进一步提高因果效应估计的精度。以下介绍两种主流方法：Gao和Ding \citep{gao2023causal}提出的加权最小二乘调整以及Lu等人\citep{lu2024adjusting}提出的网络依赖回归调整。

记在如下加权最小二乘回归$Y_i \stackrel{\omega_i}{\sim} I(\bs{D}_i=\bs{d}) + I(\bs{D}_i = \bs{d}^\prime)$，$\omega_i = 1/\pi_i(\bs{D}_i)$，中$I(\bs{D}_i=\bs{d})$和$I(\bs{D}_i = \bs{d}^\prime)$的估计系数分别为$\hat{\alpha}(\bs{d})$和$\hat{\alpha}(\bs{d}^\prime)$。Gao和Ding \citep{gao2023causal}提出的加权最小二乘调整估计为$\hat{\tau}_{\textrm{haj}} = \hat{\alpha}(\bs{d}) - \hat{\alpha}(\bs{d}^\prime)$，并证明了$\hat{\tau}_{\textrm{Haj}}$是$\tau$的相合估计且具有渐近正态性。该方法的优势在于计算过程简单，但无法保证其估计精度一定优于不调整的估计。

为了确保估计精度的提升，Lu等人\citep{lu2024adjusting}提出了依赖网络的回归调整方法。回归调整本质上是利用了$\bs{x}_i$对于$Y_i$的解释能力。Lu等人\citep{lu2024adjusting}注意到存在溢出效应时，仅依靠\(\bs{x}_i\)无法充分挖掘\(Y_i\)的解释信息，还需纳入与网络结构相关的变量。因此定义了一类新的用于回归调整的变量$\bs{G}: \mathcal{N}_n \times \mathcal{X}_n \times \{0,1\}^n\times \mathcal{E}_n\rightarrow \mathbb{R}^Q$。这里，$\bs{G}_i = \bs{G}(i,\bs{x},\bs{Z},\bs{E})$结合了协变量、处理变量以及网络的信息。在实际应用中，$\bs{G}_i$和$\bs{D}_i$的选取方式类似，可以基于一些先验的领域知识。Lu等人\citep{lu2024adjusting}提出的回归调整估计量为$\hat{\tau}(\bs{\beta}) = n^{-1}\sum_{i=1}^n w_{i,\textnormal{Lu}}(Y_i - \bs{G}_i^\top\bs{\beta}) $，其中权重
$w_{i,\textnormal{Lu}} = I(\bs{D}_i = \bs{d})/\pi_i(\bs{d})- I(\bs{D}_i = \bs{d}^\prime)/\pi_i(\bs{d}^\prime)$。不同于Gao和Ding \citep{gao2023causal}直接用加权最小二乘得到回归系数$\bs{\beta}$，Lu等人\citep{lu2024adjusting}用$\hat{\tau}(\bs{\beta})$的HAC方差估计作为损失函数，然后通过最小化该损失函数得到$\bs{\beta}$的估计。Lu等人的方法可以保证估计精度的提升。

\subsection{异质性可加效应模型}
第7.3节提到的方法通常要求网络具有一定的稀疏性。而解决网络干扰问题的另一个策略是对潜在结果施加一定的模型假设，其可以适用于更加一般的相对稠密的网络。其中具有代表性的模型是异质可加处理效应模型（heterogeneous additive treatment effect model; HATEM; \cite{sussman2017elements,harshaw2023design,yu2022estimating}）。

考虑一个$n$个个体参与的伯努利试验。我们仍然使用前文的记号。
假设网络干扰通过一个潜在的邻接矩阵$\tilde{\bs{E}} = (\tilde E_{ij})_{1\leq i,j\leq n} \in \{0,1\}^{n\times n}$来表示，其中$\tilde E_{ii}=0$且$\tilde E_{ij}=1$代表个体$j$的处理分配会影响个体$i$的潜在结果。实际问题中$\tilde{\bs{E}}$可能是未知的，但我们能观察到一个更大的网络$\bs{E}=(E_{ij})_{1\leq i,j\leq n}$，$E_{ij} \geq \tilde{E}_{ij}$。例如，Twitter上的社交网络，$E_{ij}=1$代表个体$i$关注了个体$j$，但在真实世界中真正影响个体$i$结果的可能只有其关注的个体中的一部分。记影响个体$i$的邻居数量为$\tilde{n}_i = \sum_{j=1}^n \tilde{E}_{ij}$。记观察网络的平均密度为$\rhon=\sum_{i=1}^n \sum_{j=1}^n E_{ij}/n^2$。记$\tilde{\bs{E}}$的行单位化矩阵为$\bs{Q}$，即$Q_{ij}=\tilde{E}_{ij}/\tilde{n}_i$如果$\tilde{n}_i>0$，否则$Q_{ij}=0$。

如果对网络结构和结果模型都不加任何限制，那么处理效应将会无法识别。因此，已有的关于网络干扰的研究都会对二者做一个权衡。为了得到适用于普遍网络干扰结构的结论，Lu等人\citep{lu2024estimation}考虑使用如下的异质可加处理效应模型（HATEM）:
$$Y_i(\bs{z}) = \alpha_i + \theta_i z_i + \sum_{j=1}^n \tilde{E}_{ij}\gamma_{ij} z_j.$$
其中，$\alpha_i$表示个体$i$在所有个体都分配到对照组时的基线潜在结果，$\theta_i$表示个体$i$自己的处理分配对应的直接效应，$\gamma_{ij}$表示个体$j$的处理分配对个体$i$产生的间接效应。这里，$\alpha_i,\theta_i,\gamma_{ij}$均是未知系数。在HATEM下，GATE有以下表达式:
$$\tau_{\tot} = \frac{1}{n}\sum_{i=1}^n \bigl\{Y_i(\bs{1}_n)-Y_i(\bs{0}_n)\bigr\}  = \frac{1}{n}\sum_{i=1}^n \theta_i + \frac{1}{n}\sum_{i=1}^n \sum_{j=1}^n \tilde{E}_{ij}\gamma_{ij}.$$
其中，$\bs{1}_n$是$n$维全1向量，$\bs{0}_n$是$n$维全0向量。相应地，DATE和IATE可以定义为：
$$
\tau_{\dir}  = \frac{1}{n}\sum_{i=1}^n \theta_i, \quad  \tau_{\ind}  =  \frac{1}{n}\sum_{i=1}^n\sum_{j=1}^n \tilde{E}_{ij}\gamma_{ij}.
$$
这些效应的Horvitz–Thompson估计量为
\begin{align*}
          \hat{\tau}_{\dir}   &= \frac{1}{n}\sum_{i=1}^n \Bigl\{\frac{Y_iZ_i}{r_1} - \frac{Y_i(1-Z_i)}{r_0}\Bigr\},\\
          \hat{\tau}_{\ind}  &= \frac{1}{n}\sum_{i=1}^n \sum_{j=1}^n  E_{ij} \Bigl\{\frac{Y_iZ_j}{r_1} - \frac{Y_i(1-Z_j)}{r_0}\Bigr\}, \quad  \hat{\tau}_{\tot} = \hat{\tau}_{\ind} + \hat{\tau}_{\dir}.
\end{align*}
Lu等人\citep{lu2024estimation}证明了以上Horvitz–Thompson估计量是无偏估计。为了建立这些估计量的渐近理论，Lu等人\citep{lu2024estimation}对处理效应系数以及网络结构做出一些假设：
\begin{assumption}
\label{a:bounded-parameter}
    存在常数$C$使得$\max \{\max_{1 \leq i \leq n} |\alpha_i|,\max_{1 \leq i \leq n} |\theta_i|, \max_{1 \leq i,j \leq n} \tilde{n}_i|\tilde{\gamma}_{ij}| \} \leq C$。
\end{assumption}
\begin{assumption}
\label{a:density-rho_N}
 $\rhon=o(1)$，且存在常数$C_{+}>0$使得$n\rhon\geq C_{+}$.
\end{assumption}
\begin{assumption}
\label{a:opnorm-EE^T}
     (i) $\|\bs{E}\|_{\textnormal{op}} = O(n\rhon)$且(ii) $\|\bs{Q}\|_{\textnormal{op}} = O(1)$.
\end{assumption}
直观上地，假设\ref{a:bounded-parameter}要求结果变量有界，这是网络干扰问题中的一个标准假设\citep{aronow2017estimating,leung2022causal,li2022random}。此外，也要求每个个体受其邻居影响的间接效应大小阶数一致。假设\ref{a:density-rho_N}排除了极端稠密和极端稀疏的网络。假设\ref{a:opnorm-EE^T}则是要求观察网络和真实网络中各个个体邻居数量的阶数相近。
\begin{theorem}[\citep{lu2024estimation}]
    \label{thm:var-order-of-three-estimator}
    在假设\ref{a:bounded-parameter}--\ref{a:opnorm-EE^T}下，对于$\dagger\in \{ \dir, \ind, \tot \}$，我们有$E (\hat{\tau}_{\dagger}) = {\tau}_{\dagger}$。此外，$\Var(\hat{\tau}_{\dir} ) = O(n^{-1})$，$\Var(\hat{\tau}_{\ind} )  =  O(n\rhon^2)$，$\Var(\hat{\tau}_{\tot} )  = O(n\rhon^2)$。
\end{theorem}
定理\ref{thm:var-order-of-three-estimator}表明，在HATEM下，直接效应的Horvitz–Thompson估计量始终是$n^{1/2}$相合的，而间接和全局效应估计量的相合性则需要满足$\rhon = o(n^{-1/2})$，即网络不能太稠密。为了进行有效的推断，Lu等人\citep{lu2024estimation}证明了各个估计量的渐近正态性。
\begin{theorem}[\citep{lu2024estimation}]
\label{thm:CLT-no-adjust}
    在假设\ref{a:bounded-parameter}--\ref{a:opnorm-EE^T}下，如果(i) $\max_{1\leq j\leq n}(\sum_{i=1}^n Q_{ij})^2 = o(n)$，$\max_{1\leq i\leq n} (\sum_{j=1}^n E_{ij})^2 = o(n^3\rhon^2)$，$\max_{1\leq j\leq n}(\sum_{i=1}^n E_{ij})^2 = o(n^3\rhon^2)$，(ii) $\liminf_n n\Var(\hat{\tau}_{\dir} )>0$，$\liminf_n (n\rhon^2)^{-1}$ $\Var(\hat{\tau}_{\ind} )>0$，$\liminf_n (n\rhon^2)^{-1}\Var(\hat{\tau}_{\tot} )>0$，则有
    \[
    \frac{\hat{\tau}_{\dir} -\tau_{\dir} }{\Var(\hat{\tau}_{\dir} )^{1/2}} \xrightarrow{d} \mathcal{N}(0,1),~~\frac{\hat{\tau}_{\ind} -\tau_{\ind} }{\Var(\hat{\tau}_{\ind} )^{1/2}}\xrightarrow{d} \mathcal{N}(0,1),~~\frac{\hat{\tau}_{\tot} -\tau_{\tot} }{\Var(\hat{\tau}_{\tot} )^{1/2}}\xrightarrow{d} \mathcal{N}(0,1).
    \]
\end{theorem}
值得一提的是，许多已有文献\citep{aronow2017estimating,harshaw2023design}中渐近正态性的建立都基于依赖图方法\citep{ross2011fundamentals}，这要求网络的最大度数是$o(n^{1/2})$的。而定理\ref{thm:CLT-no-adjust}对网络密度的限制仅为$\rhon=o(1)$，从而能涵盖实际应用中更多的稠密形网络。

通过理论推导，Lu等人\citep{lu2024estimation}提出了各类效应的方差估计量：
\begin{align*}
    &\hat{V}_{\dir} = \frac{1}{n^2}\Big( \sum_{i:Z_i=1} 
\frac{Y_i^2}{r_1^2}  + \sum_{i:Z_i=0} \frac{Y_i^2}{r_0^2} \Big), \quad
    \hat{V}_{\ind} = \frac{1}{n^2}\Big( \sum_{i:Z_i=1} 
\frac{A_i^2}{r_1^2}  + \sum_{i:Z_i=0} \frac{A_i^2}{r_0^2} \Big),\\
&\qquad \qquad\quad \hat{V}_{\tot} = \frac{1}{n^2}\Big\{ \sum_{i:Z_i=1} 
\frac{(Y_i+A_i)^2}{r_1^2}  + \sum_{i:Z_i=0} \frac{(Y_i+A_i)^2}{r_0^2} \Big\},
\end{align*}
其中$A_i = \sum_{j=1}^n E_{ji}Y_j$表示个体$i$的邻居结果总和。可以证明，在定理\ref{thm:CLT-no-adjust}的假设下，这些方差估计量的两倍分别为真实方差的渐近保守估计。当网络足够稠密（例如，真实网络的最小度数趋向无穷）时，这些估计量本身即可作为对应方差的保守估计。相比于文献中常用的HAC方差估计，Lu等人\citep{lu2024estimation}通过数值模拟验证了在保证第一类错误率受控的前提下，他们所提出的方差估计显著降低了保守性，从而能够构造更短的 Wald 型置信区间。

回顾定理 \ref{thm:var-order-of-three-estimator} 中间接效应和全局效应的方差阶数表达式，其主项正比于 $n^{-2} \bs{\mu}_Y^\top \bs{E} \bs{E}^\top\bs{\mu}_Y$，其中 $\bs{\mu}_Y = (\ope(Y_1), \ldots, \ope(Y_n))^\top$。为了减小该项对方差的贡献，Lu等人\citep{lu2024estimation}提出在 $\bs{E}\bs{E}^\top$ 的主要特征向量上对结果变量进行回归投影。具体而言，他们建议将个体按处理组和对照组分为两组，并将结果变量$Y_i$在 $\bs{E}\bs{E}^\top$ 的前 $K$ 个最大特征值对应的特征向量上分别进行最小二乘回归，记回归残差为 $\hat{e}_i$。对于间接效应和全局效应，可以用 $\hat{e}_i$ 替代 $Y_i$，从而构造特征向量调整后的估计量：
$$\hat{\tau}_{\ind} ^{\ev} = \frac{1}{n}\sum_{i=1}^n\sum_{j=1}^n  E_{ij} \Bigl\{\frac{\hat{e}_i Z_j}{r_1} - \frac{\hat{e}_i (1-Z_j)}{r_0}\Bigr\},\quad \hat{\tau}_{\tot} ^{\ev} = \hat{\tau}_{\dir} +\hat{\tau}_{\ind} ^{\ev}.$$
与定理 \ref{thm:CLT-no-adjust} 类似，可以证明，在一定正则条件下，特征向量调整估计量依然是渐近正态的。同时，通过在 $\hat{V}_{\ind}$ 和 $\hat{V}_{\tot}$ 中用 $\hat{e}_i$ 替代 $Y_i$，可得到特征向量调整估计量的方差估计，并可证明其保守性。与未调整的估计量相比，特征向量调整估计量具有更小的渐近方差，并在数值试验中表现出显著的精度提升。

\subsection{溢出效应和延滞效应}

在复杂的现实场景中，个体间的相互作用不仅表现为空间上的依赖性（即溢出效应），也体现在时间维度上的持续影响（即延滞效应），这两种现象在众多领域中都普遍存在。例如，在公共卫生领域，疫苗的推广应用是诠释这两种效应的典型场景：一方面，疫苗接种不仅保护个体，更能通过降低社区传播风险来惠及未接种者，形成“群体免疫”，这是一种积极的溢出效应；另一方面，个体的既往接种史对其未来的感染风险具有持续的保护作用，这正是延滞效应的体现。类似地，在教育领域，学生的学业成绩不仅受到同伴的即时影响，也深植于其过往的学习积累。实践中，溢出效应与延滞效应常常交织出现，共同对基于SUTVA的传统因果推断框架构成了严峻挑战。

近年来，部分研究开始将延滞效应纳入因果推断框架中。例如，Boruvka等人\citep{boruvka2018}聚焦于移动健康领域中时变的因果效应，在潜在结果框架内提出了新的效应定义与加权最小二乘估计方法。Bojinov 和 Shephard \citep{bojinov2019}则将潜在结果框架拓展至单一个体的时间序列试验，通过引入处理状态路径和潜在结果路径的概念，定义了新的时序因果估计量，并提出了精确的假设检验流程。Bojinov等人\citep{bojinov2023design}探讨了存在有限阶延滞效应时的最小最大最优（minimax optimal）试验设计问题，并为相应的估计量在最优设计下建立了中心极限定理。Basse等人\citep{basse2023minimax}研究了当干预效果随重复暴露而减弱（即“干预习惯化”）时序试验中的最优设计，并针对脉冲式（pulse）与楔形（wedge）两类设计，推导了能够同时估计瞬时效应和习惯化效应的最小最大最优设计。
然而，学界缺乏一个能够统一且稳健地同时处理溢出效应和延滞效应这两类违反 SUTVA 的试验设计框架。这正是本节所要探讨的问题。

为了应对同时存在溢出效应和延滞效应的复杂情况，Yu等人\citep{yu2025minimax}扩展了经典的潜在结果模型。通过引入一系列结构性假设，在保证模型可操作性的同时，尽可能地贴近现实。具体而言，假设试验中包含 $G$ 个相互独立且不重叠的试验中心，分别用 $g = 1, \ldots, G$ 表示。每个中心 $g$ 包含 $N_{[g]}$ 个参与个体，因此总体样本量为 $N = \sum_{g=1}^{G} N_{[g]}$。试验在一个离散的时间区间内进行，共有 $T$ 个时段，用 $t = 1, \ldots, T$ 表示。在最一般的情况下，个体$i$在时间$t$的潜在结果$Y_{i,t}$是整个试验期间所有个体处理分配矩阵$\bs{z}_{1:N,1:T}$的函数，即$Y_{i,t}(\bs{z}_{1:N,1:T})$ 。然而，这种形式的潜在结果数量会随着个体数$N$和时间长度$T$呈指数级增长，使其在实践中无法分析。因此，必须引入结构性假设来简化问题。Yu等人\citep{yu2025minimax}使用了三个核心假设：(1) 非预期性，即未来的处理不能影响过去的潜在结果；(2) $p$阶延滞效应，即任何个体在时间$t$的潜在结果仅依赖于过去$p+1$个时期的处理历史，而与更早期的处理无关；(3) 部分和分层干扰，即一个中心内的个体不会影响其他中心的个体以及在同一个中心内部干扰仅仅取决于该中心内的处理个体所占比例。

基于上述假设，Yu等人\citep{yu2025minimax}定义了直接效应和溢出效应。直接效应衡量的是在保持所在中心处理比例恒定的前提下，个体自身处理状态改变所带来的影响；溢出效应衡量的是在保持个体自身处理状态不变的前提下，所在中心处理比例的改变对个体结果的影响。这两种效应都可以通过Horvitz--Thompson估计量进行无偏估计。为了提高估计效率，Yu等人\citep{yu2025minimax}提出了一种两阶段的最小最大最优设计。在试验前，先选取若干决策点 $1=t_0<t_1<\ldots<t_L$，其集合记为 $\mathbb T$。处理分配仅在这些决策点上更新，在相邻决策点之间的时间区间 $[t_l, t_{l+1}-1]$ 内，各个个体的处理状态保持不变。
在第一阶段，在中心层面随机分配处理比例以确定各中心的处理强度。具体地，在每个决策点 $t_l$，为每个中心 $g$ 以0.5的概率从两个预设的处理比例 $q_1 \in (0,1)$ 和 $q_2 \in (0,1)$ 中随机抽取一个。一旦确定，该处理比例在区间 $[t_l, t_{l+1}-1]$ 内保持不变。在第二阶段，在个体层面实施具体处理分配。即在每个中心 $g$ 内，根据第一阶段确定的处理比例 $Q_{[g],t_l}$，通过完全随机化方式将 $N_{[g]}Q_{[g],t_l}$ 个个体分配到处理组，其余分配到对照组。

在这种设计下，我们不仅关注“如何”进行随机化，更关注“何时”进行随机化，即$\mathbb T$的最优选择。为解决此问题，Yu等人\citep{yu2025minimax}以最小化最差情况下的加权均方误差为目标，在理论上推导出了$\mathbb T$的最优选择，并设计了相应的多项式时间求解算法。此外，他们还发现，在特定的试验参数下，该最优方案具有一种简洁的结构化形式。在结构化的最优方案下，Yu等人\citep{yu2025minimax} 证明了 Horvitz--Thompson 估计量的相合性和渐近正态性，为构建置信区间和进行假设检验提供了坚实的理论基础。

\section{总结与展望}

本文系统回顾了基于设计的因果推断理论的主要发展脉络与最新研究进展。该理论以 Fisher 与 Neyman 的工作为奠基，强调利用随机化机制本身进行推断，从而在弱模型依赖下实现因果效应的识别与估计。本文分别综述了协变量平衡的试验设计方法（包括分层随机化、重随机化及其结合）、基于设计的统计推断方法（如 Fisher 精确检验、Neyman 渐近推断与因果自助法），以及低维与高维设置下的回归调整方法，并进一步讨论了不依从性、网络干扰等复杂情境下的理论拓展。整体来看，基于设计的因果推断已逐步从经典的随机化试验框架发展成为兼具理论严谨性与实践适用性的系统方法论。

随着机器学习和人工智能的兴起，研究者能够在更高维度、更大规模的数据中挖掘因果规律，这既带来了新的机遇，也提出了新的挑战：如何在复杂算法和非线性模型下保持推断的可解释性与稳健性。在此背景下，基于设计的因果推断进一步凸显其核心价值——以可验证的随机化机制为理论基础，在弱模型依赖下实现稳健的因果推断。未来的发展可从以下几个方面展开：其一，更复杂的试验设计与理论发展：未来试验将更加灵活与多层次，如结果适应性试验、平台试验（platform trials）、多阶段或滚动入组设计等\citep{cinelli2025challenges}；其二，更灵活的推断框架：共形推断（conformal inference）因可进行无模型的预测区间构造，被广泛用于机器学习和大语言模型的不确定性量化\citep{lei2021conformal,cherian2024large}，最近提出的基于设计的共形推断值得在因果推断中深入研究\citep{wieczorek2023design}；其三，因果机器学习：该方向已成为估计异质效应和提升效率的重要工具\citep{feuerriegel2024causal}，但其与设计框架的统一仍需理论突破；其四，应对现实试验中的复杂问题：在人类研究中，除数据缺失和不依从性外，删失、伦理约束、和中途突发事件普遍存在，基于设计的推断可通过充分考虑这些问题提升随机化试验的可信度和稳健性；其五，更复杂的时空干扰：现实干预常涉及时间与空间依赖，如政策扩散、传染病传播和社交网络效应，未来可发展空间–时间联合随机化与干扰可控设计，使设计理论适应非独立样本与动态系统；其六，因果知识的聚合与外推：Mitra等人\citep{mitra2022future}指出，应探索跨试验、跨数据源的因果知识综合，这意味着在保持随机化试验内部有效性的同时，融合真实世界数据以提升试验推断效率，或者实现设计层面的外推，提升外部有效性\citep{colnet2024causal}。总体而言，未来的基于设计的因果推断将继续以随机化为核心原则，向复杂系统、智能算法和真实世界数据扩展，在保持理论稳健性的同时，为不同领域提供更具可验证性和解释力的因果证据。

\bibliographystyle{sxjz}
\bibliography{ref}

\end{CJK}

\end{document}